\documentclass[manuscript]{aastex63}
\DeclareMathAlphabet\mathbfcal{OMS}{cmsy}{b}{n}
\usepackage{bm}

\bibpunct{(}{)}{;}{a}{}{,}

\def\setid#1,v #2 #3/#4/#5 #6 #7 #8.{(\textit{CVS version #2, checked in by #7 on #3-#4-#5, #6 UT})}

\begin{document}
\newcommand{\comment}[1]{{\color{red} #1}}
\newcommand{\mvn}[1]{{\color{magenta} michiel: #1}}
\newcommand{\fix}[1]{{#1}}
\newcommand{\lf}[1]{{#1}}
\newcommand{\lff}[1]{{#1}}
\newcommand\mymathbf[1]{{\bm{#1}}}
\title{Non-equilibrium Equation of State in stellar atmospheres}
 
\author{Anusha, L. S.$^1$, M.\ van Noort$^1$ \& R. H. Cameron}


\affiliation{Max-Planck Institute for Solar System Research,
  Justus-von-Liebig-Weg 3,
  D-37077 G\"ottingen, Germany}

\begin{abstract}
In the stellar chromospheres, radiative energy transport is dominated by only the strongest spectral lines. 
For these lines, the approximation of local thermodynamic equilibrium (LTE) 
is known to be very inaccurate, and a state of equilibrium cannot be assumed in general.
To calculate the radiative energy transport under these conditions, the population evolution equation must 
be evaluated explicitly, including all time-dependent terms.
We develop a numerical method to solve the evolution equation for the atomic-level populations in a 
time-implicit way, keeping all time-dependent terms to first order.
We show that the linear approximation of the time dependence of the populations can handle very large time steps without losing the accuracy. We reproduce the benchmark solutions 
from earlier, well-established works in terms of non-LTE kinetic equilibrium solution and typical 
ionization/recombination time-scales in the solar chromosphere.   
\end{abstract}


\section{Introduction}
\label{sec:introduction}
Since the introduction of the routine application of computing power in astrophysics, it has become an 
integral part of the interpretative process at almost every level. In particular, the evaluation of the 
expected behavior of astrophysical systems, under the assumption that they obey the laws of physics, has 
become a powerful tool in the interpretation of observational data. In solar physics, a new level of 
sophistication was made possible with the ground-breaking work of \citet[][]{1982A&A...107....1N}, 
and others, with the introduction of ab inito calculation of the solar atmosphere, where it is possible 
to work with a strongly reduced set of assumptions and approximations. The last decade in particular has 
seen a sharp rise in the use of such ``parameter-free'' simulations in the interpretation of observations, 
due to the ever-increasing availability of massively parallel computing resources.
To this end, several radiative magnetohydrodynamics (MHD) codes have been developed to simulate solar/stellar atmospheres, such as STAGGER \citep[][]{1998ApJ...499..914S}, MURaM \citep[][]{2005A&A...429..335V,2017ApJ...834...10R}, BIFROST \citep[][]{2011A&A...531A.154G},
CO5BOLD \citep[][]{2012JCoPh.231..919F}, MANCHA \citep[][]{2010ApJ...719..357F}, and so on.

The photosphere is dominated by a high gas density and relatively insignificant radiative losses, leaving the problem dominated by relatively local terms. In addition, the high density ensures high collision rates, resulting in time scales on which the atomic populations reach an equilibrium with their environment that is very short compared to the time scale on which that environment changes, so that equilibrium conditions may safely be assumed. This condition is known as the local thermodynamic equilibrium (LTE). Due to the low particle densities, however, this is no longer the case in the chromosphere, where the collisional rates are low and processes involving multiple particles are correspondingly rare. The evolution time scale of the magnetically dominated structures, on the other hand, is considerably shorter than in the photosphere, so that the time scales on which the atomic populations reach an equilibrium with their environment may well exceed the evolution time scale of the environment itself, so that equilibrium 
is never reached. 

\citet[][]{1976ApJ...205..499K,1978ApJ...220.1024K} and \citet[][]{1980A&A....87..229K} demonstrated that the assumption of kinetic (statistical) equilibrium produces contrasting ionization and recombination time scales, thus proving the invalidity of that assumption in a dynamic atmosphere with shock waves. \citet[][]{2017ApJ...851....5J} proposed a fast probabilistic approach to solve the non-LTE non-equilibrium radiative transfer (RT) for dynamically evolving one-dimensional (1D) atmospheres. The RADYN code \citep[][]{1992ApJ...397L..59C,1995ApJ...440L..29C,2002ApJ...572..626C} solves the 1D hydrodynamic equations (equation of mass, momentum and energy conservation) together with the non-LTE RT equation using an adaptive mesh algorithm. Although RADYN takes care of the non-equilibrium and non-LTE effects consistently, because it is a a hydrodynamical code it cannot be used to study heating mechanisms that rely on the presence of magnetic fields. For these reasons, RADYN is not suitable for three-dimensional (3D) radiative MHD simulations. 

The BIFROST code \citep[][]{2011A&A...531A.154G} is capable of radiative MHD simulations in two dimensions (2D) and 3D that also takes nonequilibrium effects into account \citep[][]{2007A&A...473..625L,2009ApJ...694L.128L}. However, in BIFROST, the time dependence is neglected in the rate system, and it further approximates equation-of-state and RT calculations by using prescribed recipes for chromospheric radiative losses and hydrogen ionization in order to make the problem more tractable in 2D/3D. 

Generally, in the radiation MHD simulations, chemical equilibrium is assumed for the molecular formation/dissociation in the equation of state.
Although a simple nonequilibrium treatment of the ${\rm {H_2}}$ molecule is included in \citet[][]{2011A&A...530A.124L}, there is potential for studying the importance of other chemical reactions in the solar atmosphere.

In this paper, we focus on improving the existing methods to solve the non-LTE nonequilibrium RT problem through (a) a proper time-dependent treatment of the radiation
field,
(b) a proper nonequilibrium treatment of the molecular chemistry, and
(c) development of a time-implicit numerical scheme.
Furthermore, the method developed in this paper is suitable for 1D/2D/3D simulations.

In Section~\ref{sec:rmhd} we describe basic radiative MHD, RT, and kinetic equilibrium equations. In Section~\ref{sec:theory} we describe in detail our development of the population evolution scheme, our newly developed time-dependent short-characteristics method for the formal solution of the RT equation, and finally, our generalization of the multi-level approximate lambda iteration (MALI) scheme to solve the time-dependent nonlinear rate system. In Section~\ref{sec:implementaion} we describe the implementation of the method. In Section~\ref{sec:verification} we discuss the accuracy of our new method and show the benchmark solutions reproduced from our method. Finally, we present a summary in Section~\ref{sec:summary}.

\section{Radiation MHD}
\label{sec:rmhd}
\subsection{Basic equations}
The MHD equations are typically expressed as conservation equations, 
expressing the conservation of mass, momentum, energy, and magnetic flux along with a number of expressions to close the system.

We recall the important hydrodynamic equations here, ignoring the magnetic field for the sake of simplicity. 
The equations of continuity, momentum, and energy are respectively given by 
\citep[see, e.g.,][]{2014ApJ...789..132R}
\begin{eqnarray} \label{eq:hdeq}
\frac{\partial \rho}{\partial \,t}=\nabla \cdot (\rho \bm{v}),\nonumber \\
\frac{\partial \rho \bm{v}}{\partial \,t}=-\nabla \cdot (\rho \bm{v}\bm{v}) -\nabla P+\rho \bm{g},
\nonumber \\
\frac{\partial E_{\rm {HD}}}{\partial \,t} = -\nabla \cdot [\bm{v} (E_{\rm {HD}}+ P)]+\rho \bm{v} \cdot 
\bm{g} + Q_{\rm {rad}}.
\end{eqnarray}

Here $\rho$, $P$ and $\bm{v}$ denote the mass density, pressure, and velocity, respectively. We note that the atmosphere we are considering is a small box encompassing part of the convection zone, the photosphere, chromosphere, and part of the corona. Therefore, we approximate the gravitational acceleration to be a constant taken to be $-2.74 \times 10^4 {\rm{cm^2 s^{-1}}}$. In the solar/stellar chromosphere, the dominant term in the energy equation is the 
loss of energy to the outer environment in the form of radiation. Here $Q_{\rm {rad}}$ is the radiative heating term given by the divergence of radiative flux $-\nabla \cdot F$,
where $F$ is the radiative flux. 

\begin{figure*}
\includegraphics[scale=1]{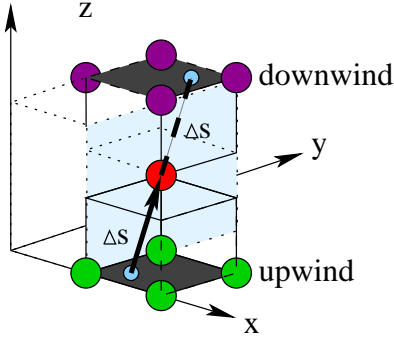}
\caption{Short characteristics on a 3D grid.}
\label{fig:shortchar}
\end{figure*}

\subsection{Equations of state} 
\label{eos}
We assume that we have a gas mixture that consists of atomic hydrogen, denoted by ${\rm{H}}$; the hydrogen molecules ${\rm{H_2}}$, ${\rm{H_2^+}}$ and ${\rm{H^-}}$; and several other metals.
The occupation numbers of the constituents of the gas determine the pressure $P$ and 
$F$ is computed using the specific intensity obtained by solving the RT equation. When the approximation of LTE is valid,
the solution of the RT equation is simple, and the pressure can be tabulated. For non-LTE conditions, such as the solar/stellar chromosphere, one cannot use tables, and explicit treatment of
the elements of the gas needs to be considered. 

The total energy is distributed into the kinetic and internal energies of the constituents of the gas. The internal energy of the system is distributed amongst its various degrees of freedom (e.g., bound and free states of atoms and molecules). The temperature can be determined by imposing the total energy to be conserved. Finally, the system is closed by imposing particle and charge conservation.
The combinations of these expressions are known as the equations of state. 

\subsection{The RT equation}
The flux divergence term $Q_{\rm {rad}}=\nabla \cdot F$ appearing in the energy equation (see Equation~\ref{eq:hdeq}) requires the radiative flux $F$ which is given by
\begin{equation}
F=\int d\nu \oint d {\bm \Omega}\,\,\,\,{\bm \Omega} \cdot {I}_{\nu,{\bm \Omega}}, \end{equation}
where ${I}_{\nu,{\bm \Omega}}$ is the specific intensity at frequency $\nu$ along the ray direction ${\bm \Omega}$ \citep[see, e.g.,][]{1978stat.book.....M,2014tsa..book.....H}. To obtain ${I}_{\nu,{\bm \Omega}}$ we need to solve the RT equation, which, in Cartesian geometry, is given by
\begin{eqnarray}
-\frac{1}{\alpha_{\rm {tot}}}{\bm{\Omega}} \cdot \nabla {I}_{\nu,{\bm \Omega}}={I}_{\nu,{\bm \Omega}}-{S}_{\nu},
\end{eqnarray}
where ${S}_{\nu}=\frac{\eta_{\rm {tot},\nu}}{\alpha_{\rm{tot}},\nu}$ is the source function, with $\eta_{\rm{tot},\nu}$ and $\alpha_{\rm{tot},\nu}$ being the total emissivity and total opacity, taking contributions from both the lines and the continua. The emissivity and opacity depend on the atomic population density ${\bm {n}}$. The solution of the RT equation, known as the formal solution, depends nonlinearly on ${\bm {n}}$.

\noindent
We note here that we ignore the term $\frac{1}{c} \frac{d I_{\nu}}{dt}$ in the RT equation.  As studied in detail in \citet[][]{1976ApJ...205..499K}
the ratio of the thermal relaxation time (the time needed to lose the internal energy through radiation for a heated gas) to the photon travel time is $\sim$ $10$-$10^3$; also, the photon travel time is much shorter when compared with the dynamic time-scales (which are of the order of scale height to local sound speed $\sim$ 10). This means that we can safely assume that the radiation field adjusts instantaneously to the state of the medium. 

\subsection{Kinetic equilibrium equation}
\label{see}
In general, we will assume that the rate of change of the population densities ${\bm n}$ of atomic excited states in a medium is given by the coupled system of equations
\begin{equation}
\frac{\partial {\bm n}}{\partial t}+({\bm v}\cdot\nabla) {\bm n}={\bm A}\cdot{\bm n},
\label{eq:popevol}
\end{equation}
where the rate matrix ${\bm A}$ typically contains the radiative and collisional transition rates between the different atomic energy levels. The radiative rate coefficients are themselves nonlinearly dependent on the instantaneous populations, so that the system of equations does not have an analytical solution in general. In many situations of interest, stationary or quasi-stationary conditions can be assumed, implying that
\begin{equation}
\frac{\partial {\bm n}}{\partial t}= {\bm 0},
\,\,\,\,\,(\bf{v} \cdot \nabla){\bm n} = {\bm 0},
\nonumber \\
\end{equation}
and thus the populations must satisfy the condition 
\begin{equation}
{\bm A}\cdot{\bm n}=0,
\nonumber \\
\end{equation}
or, in a component form,
\begin{equation}
\sum_{l^\prime\ne l} n_{l^\prime} C_{l^\prime,l} + \sum_{l^\prime> l}n_{l^\prime} R_{l^\prime,l} - \sum_{l^\prime\ne l}n_l  C_{l,l^\prime} -\sum_{l^\prime< l}n_l R_{l,l^\prime} = 0,
\label{eq:see}
\end{equation}
where $l$ is the level index of the population densities. Besides the locally determined collisional rates $C_{l,l^\prime}$ these equations contain radiative rates $R_{l,l^\prime}$ that depend on the angular and transition -averaged intensity $\bar{J}_{l^\prime,l}$, which is highly non-local in optically thin conditions. Equation ~(\ref{eq:see}) is known as the {\it kinetic equilibrium equation} or {\it statistical equilibrium equation}. The solution of this system of equations is typically found by iterative evaluation of the 
RT equation in order to improve an initial estimate of ${\bm n}$, 
a process that is usually referred to as non-LTE RT.

\subsection{Time-scales}
\label{timescales}
An important aspect of the non-LTE radiation field in the solar/stellar chromosphere is that the time scale of the radiative recombination process can vary from $\sim$ 50 up to $10^5$ s \citep[][]{1992ApJ...397L..59C,1995ApJ...440L..29C,2002ApJ...572..626C}, which is much longer than the dynamic time-scale of $\sim$ $10$ s  \citep[][]{1976ApJ...205..499K}. Therefore, the assumption of instantaneous kinetic (statistical) equilibrium (Equation~\ref{eq:see}) is not always valid in the chromosphere. Therefore, we need to treat the atomic populations in non-equilibrium. Further, due to the interdependence of the of non-LTE radiation field and the atomic populations, we need to solve the non-LTE RT equation and the nonequilibrium rate-system, consisting of collisional and radiative rates, to obtain the occupation numbers of all of the constituent atomic species simultaneously.

\section{Evolution of state}
\label{sec:theory}
\subsection{Population}
\label{sec:popevolution}
We have seen that under the kinetic equilibrium conditions, we have Equation~(\ref{eq:see}), the solution of which, together with the simultaneous solution of the RT equation, provides the population density ${\bm n}$. As discussed in Section~\ref{timescales} we cannot assume kinetic equilibrium in general. When equilibrium cannot be assumed, we are left with no choice but to evaluate Equation~(\ref{eq:popevol}), which has the form
\begin{equation}
{\bm A}\cdot{\bm n}={\bm b},
\nonumber \\
\label{eq:ne-system}
\end{equation}
or, in component form with level index $l$,
\begin{equation}
\frac{\partial n_l}{\partial t}+({\bm v}\cdot\nabla) n_l=\sum_{l^\prime\ne l} n_{l^\prime} C_{l^\prime,l} + \sum_{l^\prime> l}n_{l^\prime} R_{l^\prime,l} - \sum_{l^\prime\ne l}n_l  C_{l,l^\prime} -\sum_{l^\prime< l}n_l R_{l,l^\prime}.
\label{eq:rates}
\end{equation}
We note here that Equation~(\ref{eq:rates}) has the same form as that of Equation~(\ref{eq:see}), which is the non-LTE kinetic equilibrium equation. However, we now have a nonzero vector ${\bm b}$ on the right-hand side. We solve this system by integrating Equation~(\ref{eq:popevol}) from time $t_n$ to time $t$, resulting in the formal solution
\begin{equation}
{\bm n}(t)={\bm n}(t_n) + \int_{t_n}^{t}{\bm A}(t^\prime)\cdot{\bm n}(t^\prime)d\,t^\prime-\int_{t_n}^{t}({\bm v}(t^\prime)\cdot\nabla){\bm n}(t^\prime) d\,t^\prime.
\label{eq:formalsol}
\end{equation}
To find a solution to this system,
we follow the ideas that lead to the short-characteristics method of \citet[][]{1988JQSRT..39...67K} which was originally developed for an efficient evaluation of the formal solution of the non-LTE RT equation in a multidimensional geometry. Along each ray of the angle quadrature direction, instead of traversing the entire space, the intensity is evaluated locally at the central point of a three-point stencil, using the known intensity in the upwind direction. The nonlinear spatial dependence of the source function is approximated using a polynomial in terms of the known source function values at the spatial grid points on the local stencil. The polynomial form of the source function allows a direct evaluation of the formal integral. This process is repeated along the ray to cover the entire space \citep[see also][]{1994A&A...285..675A}.

We now apply the same ideas to the time variable $t$. Denoting $t$ at two successive time steps as $t_n$ and $t_{n+1}$, we define $\delta\,t=t_{n+1}-t_{n}$, as the time interval between $t_n$ and $t_{n+1}$.
We are interested in evaluating ${\bm n}(t)$ at $\delta\,t$ for each time step. For this purpose, all of the time-dependent quantities are expressed as linear polynomials in $t$ that use the known values of these quantities at previous time step and ignore all of the cross-terms of second and higher order.  This allows a direct evaluation of the time integration in Equation~(\ref{eq:formalsol}), yielding a linear system of equations. \\

We start with the linear polynomial form of the populations,
\begin{equation}
{\bm n}_n(t)={\bm n}_n+\dot{{\bm n}}_n (t-t_n),
\nonumber \\
\end{equation}
and substitute it in Equation~(\ref{eq:formalsol}), which yields
\begin{equation}
\dot{\bm n}_n t  = \int_{t_n}^{t}{\bm A}(t^\prime)\cdot\left( \dot{\bm n}_n t^\prime+{\bm n}_n\right) d\,t^\prime-\int_{t_n}^{t}({\bm v}(t^\prime)\cdot\nabla)\left(\dot{\bm n}_n t^\prime+{\bm n}_n\right)  d\,t^\prime.
\label{eq:newformalt}
\end{equation}

At $\delta\,t$, we have 
\begin{equation}
\dot{\bm n}_n \delta\,t   = \int_{0}^{\delta\,t}{\bm A}(t^\prime)\cdot\left(\dot{\bm n}_n t^\prime+{\bm n}_n\right) d\,t^\prime-\int_{0}^{\delta\,t}({\bm v}(t^\prime)\cdot\nabla)\left(\dot{\bm n}_n t^\prime+{\bm n}_n\right)  d\,t^\prime,
\label{eq:newformal}
\end{equation}
which can be solved for $\dot{\bm n}_n$.
The main difficulty in solving this system lies in the implicit nonlocal and nonlinear dependence of 
the first term on the right-hand side on $\dot{\bm n}_n$. This dependence is similar to that found in equilibrium 
non-LTE problems, for which it is known that it is very stiff and converges very slowly when 
solved by iterative means only. 
A more successful approach is to linearize and 
localize Equation~(\ref{eq:newformal}) in $\dot{\bm n}_l$ and solve the linear system, while solving for the 
remaining nonlinearity iteratively, a process that is usually referred to as ``acceleration''  \citep[see][and the references cited therein]{1973ApJ...185..621C,1973JQSRT..13..627C,1981ApJ...249..720S,1986JQSRT..35..431O,1991A&A...245..171R,1992A&A...262..209R,2003ASPC..288...17H}.

We therefore focus our attention on the nonlinear term and expand it,
\begin{equation}
\int_{0}^{\delta\,t}{\bm A}(t^\prime)\cdot\left(\dot{\bm n}_n t^\prime+{\bm n}_n\right) d\,t^\prime
= \int_{0}^{\delta\,t}{\bm A}(t^\prime) t^\prime d\,t^\prime \cdot \dot{\bm n}_n +\int_{0}^{\delta\,t}{\bm A}(t^\prime) d\,t^\prime \cdot {\bm n}_n.
\nonumber \\
\end{equation}
In light of Equation~(\ref{eq:rates}), this expression contains three types of integrals over time that must be calculated. The integrals over the constant spontaneous emission terms are trivial and is not discussed. The collisional rates, however, are a complex function of the temperature $T_n(t)$ of the form
\begin{equation}
C_{l,l^\prime}=C_0 n_e(t) \sqrt{T(t)}\,e^{-\frac{E_0}{k_B T(t)}} \Gamma[T(t)].
\nonumber \\
\end{equation}
Assuming the time dependence of the temperature, the electron density, and the coefficients $\Gamma[T(t)]$ 
in the time interval $[t_n,t_{n+1}]$ to be linear,
\begin{equation}
{T}(t)=\dot{{ T}}_n t+{ T}_n,
\nonumber \\
\end{equation}
\begin{equation}
n_e(t)=\dot{n}_{{e,}_n} t+n_{{e,}_n},
\nonumber \\
\end{equation}
\begin{equation}
\Gamma(t)=\dot{\Gamma}_n t+{\Gamma}_n,
\end{equation}
we can 
express the collisional rate coefficients as linear polynomials in $t$ and write them as
\begin{equation}
C_{l,l^\prime}=\dot{C}_{l,l^\prime,n} t+{C_{l,l^\prime,n}}.
\end{equation}

The time-integrated radiative rates are given by
\begin{eqnarray}
&& \int_{0}^{\delta\,t} R_{l,l^\prime}(t^\prime) d\,t^\prime= \int  d\,t^\prime \oint  d\,\bm{\Omega} \frac{ d\,\nu}{h\,\nu}
[U_{l,l^{\prime}}(t^\prime)+n_e(t^\prime) U_{l,l^{\prime}}^{\star}(t^\prime)+
  (V_{l,l^{\prime}}(t^\prime)+n_e(t^\prime) V_{l,l^{\prime}}^{\star}(t^\prime)) I(t^\prime)],
\nonumber \\ 
\end{eqnarray}
where the quantities $U_{l,l^{\prime}}$, $U_{l,l^{\prime}}^{\star}$, $ V_{l,l^{\prime}}$ and $V_{l,l^{\prime}}^{\star}$ are defined in Appendix~\ref{appendixa}.
Evaluation of this expression presents a challenge, however, since the intensity is a highly nonlinear, 
nonlocal function of the populations and, is in addition, a complicated function of time. Upon 
substituting the time-linear expression for all of the time-dependent quantities (see Appendix~\ref{appendixa} for details) and neglecting all terms of second or higher order in time in the above equation, we have
\begin{eqnarray}
\int_{0}^{\delta\,t} R_{l,l^\prime}(t^\prime) d\,t^\prime &=&  \oint  d\,\bm{\Omega} \frac{ d\,\nu}{h\,\nu}
[(U_{l,l^{\prime}}+n_e U_{l,l^{\prime}}^{\star}) \delta t+
(\dot{U}_{l,l^{\prime}}+n_e \dot{U}_{l,l^{\prime}}^{\star}+ \dot{n}_e U_{l,l^{\prime}}^{\star}) \frac{(\delta t)^2}{2}
\nonumber \\
  &&+(V_{l,l^{\prime}}+n_e V_{l,l^{\prime}}^{\star}) \bar{I}+
(\dot{V}_{l,l^{\prime}}+n_e \dot{V}_{l,l^{\prime}}^{\star}+ \dot{n}_e V_{l,l^{\prime}}^{\star}) \hat{I}
  ],
\nonumber \\ 
\end{eqnarray}
where
\begin{equation}
\overline{I}=\int_{0}^{\delta\,t} I(t^\prime) \, d\,t^\prime
\,\,\,\,\,\,\,\,\,\,{\rm and}\,\,\,\,\,\,\,\,\,\,
\hat{I}=\int_{0}^{\delta\,t} I(t^\prime) t^\prime \, d\,t^\prime
\label{eq:intint}
\end{equation}
were introduced.

\subsection{Radiative quantities}
To obtain the intensity at a given frequency and angle in every point on the grid, we must first be able to calculate the opacity and emissivity. The opacity at a given frequency $\nu$ is the result of the sum over all lines, added to the continuum opacity $\alpha_c$,
\begin{equation}
\alpha_\nu=\alpha_{c,\nu}+\frac{h\nu}{4\pi}\sum_{l<l^\prime} \varphi_{l,l^\prime,\nu}[B_{l,l^\prime} n_{l}-B_{l^\prime,l}n_{l^\prime}].
\label{eq:totopac}
\end{equation}
To obtain a time-dependent opacity in the interval $t=t^n$ to $t^{n+1}$ that we can work with, we must approximate the time dependence of the line profile $\varphi_{l,l^\prime}(t)$, as was done earlier for the populations $n(t)$. Under the assumption that the acceleration over each time interval is small compared to the Doppler width of the line, we can make a linear expansion of $\varphi_{l,l^\prime}(t)$ and the bulk velocity ${v}(t)$ around $t=t^n$
\begin{equation}
\varphi_{l,l^\prime}(t)\approx\varphi_{l,l^\prime}^n+  t\,\dot{\rm v}\frac{\nu}{c}\left[\frac{\partial \varphi_{l,l^\prime}(\nu^\prime)}{\partial \nu^\prime}\right]_{\nu^\prime=\nu}\equiv\varphi_{l,l^\prime}+  t\,\dot{\rm v}\frac{\nu}{c}\hat{\varphi}_{l,l^\prime},
\nonumber \\
\end{equation}
and substitute it in Equation~(\ref{eq:totopac}). The resulting opacity for the transition $l\rightarrow l^\prime$,
{\small
\begin{equation}
\alpha_{l,l^\prime}(t)=\frac{h\nu}{4\pi}(\varphi_{l,l^\prime}+ \hat{\varphi}_{l,l^\prime}\frac{\nu}{c}\dot{\rm v} t)((B_{l,l^\prime}\dot{n}_l -B_{l^\prime,l}\dot{n}_{l^\prime}) t + B_{l,l^\prime}n_l - B_{l^\prime,l}n_{l^\prime}),
\label{eq:linopac}
\end{equation}
}
is a quadratic function of time, due to the interplay between the time dependence of the population densities and the Doppler-shifted line profile. However, since we neglected all time dependence of second and higher order in the populations and in the line profile, we may, without loss of accuracy, proceed by dropping all terms of second and higher order in time from Equation~(\ref{eq:linopac}), yielding
\begin{equation}
\begin{array}{r c l}
\alpha_\nu(t)&\approx&\alpha_{c,\nu}(t)+\frac{h\nu}{4\pi}\sum_{l<l^\prime} \Big[(\varphi_{l,l^\prime} (B_{l,l^\prime}\dot{n}_l - B_{l^\prime,l}\dot{n}_{l^\prime}) + \\
&&\hat{\varphi}_{l,l^\prime}\frac{\nu}{c} \dot{\rm v} (B_{l,l^\prime}n_l - B_{l^\prime,l}n_{l^\prime}) ) t +\varphi_{l,l^\prime}(B_{l,l^\prime}n_l - B_{l^\prime,l}n_{l^\prime})\Big].
\end{array}
\nonumber \\
\end{equation}
Similarly, we approximate the emissivity
\begin{equation}
\eta(t)\approx\eta_{c,\nu}(t) +\frac{h\nu}{4\pi}\sum_{l>l^\prime} A_{l,l^\prime}\left[(\varphi_{l,l^\prime} \dot{n}_l + \hat{\varphi}_{l,l^\prime}\frac{\nu}{c} \dot{\rm v} n_l ) t +\varphi_{l,l^\prime} n_l \right].
\nonumber \\
\end{equation}
For strong spectral lines, the contribution of continuum sources to the emissivity and opacity is generally small, so that the order of the contribution to the time dependence of them can safely be assumed to be linear. Clearly, this assumption reduces the time dependence of the opacity and emissivity to linear,
\begin{equation}
\begin{array}{c}
\eta_\nu(t)\approx\dot{\eta}_\nu t +\eta_{\nu,0},\\
\alpha_\nu(t)\approx\dot{\alpha}_\nu t +\alpha_{\nu,0},
\end{array}
\label{eq:approxemop}
\end{equation}
where
{\small
\begin{equation}
\begin{array}{l}
\dot{\eta}_\nu=\dot{\eta}_{c,\nu}+\frac{h\nu}{4\pi}\sum_{l>l^\prime} A_{l,l^\prime}\left[\varphi_{l,l^\prime}
\dot{n}_l + \hat{\varphi}_{l,l^\prime}\frac{\nu}{c} \dot{\rm v} n_l \right],\\
\eta_\nu=\eta_{c,\nu}+\frac{h\nu}{4\pi}\sum_{l>l^\prime} A_{l,l^\prime} \varphi_{l,l^\prime} n_l, \\
\dot{\alpha}_\nu=\dot{\alpha}_{c,\nu}+\frac{h\nu}{4\pi}\sum_{l<l^\prime} \left[(\varphi_{l,l^\prime} (B_{l,l^\prime}\dot{n}_l - B_{l^\prime,l}\dot{n}_{l^\prime}) +
\hat{\varphi}_{l,l^\prime}\frac{\nu}{c} \dot{\rm v} (B_{l,l^\prime}n_l - B_{l^\prime,l}n_{l^\prime})\right],\\
\alpha_\nu= \alpha_{c,\nu}+\frac{h\nu}{4\pi}\sum_{l<l^\prime} \varphi_{l,l^\prime} (B_{l,l^\prime}n_l - B_{l^\prime,l}n_{l^\prime}).
\end{array}
\nonumber \\
\end{equation}
}
The source function now assumes the simple rational form
\begin{equation}
S_\nu(t)=\frac{\eta(t)}{\alpha(t)}=\frac{\dot{\eta}_\nu t +\eta_{\nu}}{\dot{\alpha}_\nu t +\alpha_{\nu}}.
\nonumber \\
\end{equation}

\subsection{Time-dependent short characteristics}
The method of the short characteristics \citep[][]{1988JQSRT..39...67K,1994A&A...285..675A} solves the RT equation along the characteristics of the equation (rays) that are limited to individual grid cells. While this method 
generates the problem of required knowledge of the intensity on the upstream boundary of each grid cell, 
an especially complicating factor when the calculations need to be parallelized, it involves 
numerical work that is constant for each grid cell and thus scales linearly with the 
number of grid cells, a property that, despite this drawback, has made it the method of choice for multidimensional problems in the last two decades.

We thus proceed by calculating the intensity on the grid using this method and formulate the integration over the optical depth in each cell using quadratic interpolation for the atmospheric properties as a function of optical depth \citep[see also][]{1988JQSRT..39...67K,1994A&A...285..675A}. We note that although the formulation provided here is for quadratic polynomials, in practice, we use linear polynomials, as they are more stable and faster to compute. Alternatively, Bezier or Hermitian polynomials could be used to improve the stability and accuracy of the method \citep[see, e.g.,][]{2003ASPC..288....3A}.

We adopt the method of the locally comoving laboratory frame \citep[see, e.g.,][]{2002ApJ...568.1066V,2009ApJ...694L.128L}, which assumes for each grid point that it is at rest. This method provides a convenient mix of the observer's frame and the comoving frame formalisms, but requires all local quantities (opacity, emissivity) to be interpolated to the locally comoving frame. A key advantage of this method is that the angle-dependent frequency redistribution of the radiation field induced by spatial gradients in the bulk velocity of the gas is automatically taken into account. In the current work, we will limit ourselves to situations where the velocity gradients are sufficiently small that the Doppler shift within one grid cell does not significantly exceed the frequency grid spacing, so that additional subgridding of the characteristics can be omitted. 

To calculate the comoving observer's frame optical depth at frequency $\nu$, we need to integrate the opacity, given by Equation~(\ref{eq:totopac}). To integrate this expression, the spatial dependence of the $\varphi_{ij,\nu}$ and $n_i$ must be known analytically. To obtain this dependence, we interpolate the relevant physical quantities from the grid to the end points of the characteristic at distances $\pm\Delta s$ from the center. To guarantee positivity, we require the analytic continuation along the characteristic to be linear in the spatial coordinate, $s$,
\begin{equation}
\alpha_\nu(s,t)\approx \alpha_\nu^c(t)+\frac{1}{\Delta s}\alpha^\prime_\nu (t) s,
\nonumber \\
\end{equation}
where $\alpha_\nu^c(t)$ is the opacity in the central grid point, and $\alpha^\prime_\nu (t)$ is the spatial derivative of the opacity, obtained by means of spatial interpolation at the intersection of the characteristic and the upwind cell boundary,
\begin{equation}
\alpha^\prime_\nu (t)=\frac{\sum_j c_j \alpha_{\nu^\prime}^{u,j}(t)}{\sum_j c_j}-\alpha_\nu^c(t),
\nonumber \\
\end{equation}
where $\alpha_\nu^{c}(t)$ and $\alpha_{\nu^\prime}^{u,j}(t)$ are the opacity values in the current grid point and the upwind grid points with interpolation index $j$ and at the Doppler-shifted frequency,
\begin{equation}
\nu^\prime(t)=\nu+\frac{\nu}{c}\left(\frac{\sum_j c_j {\rm v}^{u,j}(t)}{\sum_j c_j}-{\rm v}^c(t)\right),
\nonumber \\
\end{equation}
at the upwind limit of the characteristic respectively. The number of contributions to the sum depends on the order chosen for the interpolation and is four for the linear scheme illustrated in Fig.~\ref{fig:shortchar}. The trivial integration along the characteristic now results in the optical depth
\begin{equation}
\tau^u_\nu(t)=\int_0^{\Delta s_u} \alpha_\nu(s,t)  d\,s\approx [\alpha_\nu^c(t)+\frac{1}{2}\alpha^\prime_{\nu^\prime} (t)] \Delta s,
\nonumber \\
\end{equation}
which is readily calculated using Equation~(\ref{eq:approxemop}) to have the form
\begin{equation}
\tau^u(t)=\dot{\tau}^u t + \tau^u,
\label{eq:upwind_opticaldepth}
\end{equation}
where
\begin{equation}
\begin{array}{l}
\dot{\tau}^u=\left[\dot{\alpha}_\nu^c+\frac{1}{2} \left\{\frac{\sum_j c_j \dot{\alpha}_{\nu^\prime}^{u,j}}{\sum_j c_j}-\dot{\alpha}_\nu^c \right\}\right]\Delta s,\\
\tau^u=[\alpha_\nu^c+\frac{1}{2}\left\{\frac{\sum_j c_j \alpha_{\nu^\prime}^{u,j}}{\sum_j c_j}-\alpha_\nu^c  \right\}] \Delta s.
\end{array}
\end{equation}
Similarly, the downwind optical depth $\tau^d(t)$ can be calculated from
\begin{equation}
\tau^d(t)=\dot{\tau}^d t + \tau^d,
\label{eq:downwind_opticaldepth}
\end{equation}
where
\begin{equation}
\begin{array}{l}
\dot{\tau}^d=\left[\dot{\alpha}_\nu^c+\frac{1}{2} \left\{\frac{\sum_j c_j \dot{\alpha}_{\nu^\prime}^{d,j}}{\sum_j c_j}-\dot{\alpha}_\nu^c \right\}\right]\Delta s,\\
\tau^d=[\alpha_\nu^c+\frac{1}{2}\left\{\frac{\sum_j c_j \alpha_{\nu^\prime}^{d,j}}{\sum_j c_j}-\alpha_\nu^c  \right\}] \Delta s.
\end{array}
\end{equation}
We are now ready to evaluate the formal integral along the short characteristic. Using the method from \citet[][]{1994A&A...285..675A}, we write the source function as a parabolic function of the optical depth,
\begin{equation}
S(\tau,t)=c_0(t)+c_l(t) \tau(t) + c_q(t) \tau^2(t),
\nonumber \\
\end{equation}

and evaluate the formal integral,
\begin{equation}
I(t)=I_0(t)e^{-\tau^u(t)}+\int_0^{\tau^u(t)} \left({\rm c}_0(t) 
+ {\rm c}_l(t) s + {\rm c}_q(t) s^2\right) e^{-s}  d\, s,
\nonumber \\
\end{equation}
along the characteristic, which has the general solution
\begin{equation}
\begin{array}{r l}
I(t)=[&I_0(t)e^{-\tau^u(t)}+\\
&c_0(t)(1-e^{-\tau^u(t)})+ \\
&c_l(t)(1-(\tau^u(t)+1) e^{-\tau^u(t)}) + \\
&c_q(t)(2-(2(1+\tau^u(t))+\tau^u(t)^2) e^{-\tau^u(t)})],
\end{array}
\label{eq:formalsolution}
\end{equation}
where the coefficients of $S$ are given by
\begin{equation}
{\rm c}_0(t)=S^c(t),
\nonumber \\
\end{equation}
\begin{equation}
{\rm c}_l(t)=\frac{[S^c(t)-S^d(t)]{\tau^u(t)}^2-[S^c(t)-S^u(t)]{\tau^d(t)}^2}{\tau^u(t)\tau^d(t)[\tau^u(t)+\tau^d(t)]}\, ,
\nonumber \\
\end{equation}
and
\begin{equation}
{\rm c}_q(t)=\frac{[S^d(t)-S^c(t)]\tau^u(t)+[S^u(t)-S^c(t)]\tau^d(t)}{\tau^u(t)\tau^d(t)
[\tau^u(t)+\tau^d(t)]}\,.
\nonumber \\
\end{equation}

To obtain the time-integrated rates, we still need to integrate over frequency, angle, and time. The unappealing prospect of carrying the analytic form of Equation~(\ref{eq:formalsolution}) through to the end suggests that it might be advantageous to carry out the time integration of Equation~(\ref{eq:formalsolution}) over the time step before the angular and frequency integration.

Substitution for $S$ in the coefficients yields
\begin{equation}
{\rm c}_0=
\frac{\eta_c}{\alpha_c},
\label{eq:constcoef}
\end{equation}
\begin{equation}
{\rm c}_l=
\frac{(\alpha_u\alpha_d\eta_c-\alpha_u\alpha_c\eta_d)\tau_u^2-(\alpha_u\alpha_d\eta_c-\alpha_c\alpha_d\eta_u)\tau_d^2}{(\tau_u+\tau_d)\tau_d\tau_u\alpha_u\alpha_c\alpha_d}\, ,
\label{eq:lincoef}
\end{equation}
and
\begin{equation}
{\rm c}_q=\frac{(\alpha_u\alpha_c\eta_d-\alpha_u\alpha_d\eta_c)\tau_u+
(\alpha_c\alpha_d\eta_u-\alpha_u\alpha_d\eta_c)\tau_d}{(\tau_u+\tau_d)\tau_d\tau_u\alpha_u\alpha_c\alpha_d}\,.
\label{eq:quadcoef}
\end{equation}
in which the explicit time dependence was dropped for notational brevity. Despite the substantial simplifications made earlier, the result has clearly regained a complicated time dependence through the high-order rational form of the coefficients. 

The form of Equation~(\ref{eq:formalsolution}) suggests that it is advantageous to rewrite it to

\begin{equation}
\begin{array}{r l}
I(t)=&I_0(t)e^{-\tau^u(t)}\\
&-[c_0(t)+c_l(t)(1+\tau^u(t))+c_q(t)(2(1+\tau^u(t))+\tau^u(t)^2)]e^{-\tau^u(t)}\\
&+c_0(t)+c_l(t) + 2 c_q(t),
\end{array}
\label{eq:formalsolutionreorg}
\end{equation}
then expand and add up the coefficients $c_0$, $c_l$ and $c_q$. The resulting expression is a high-order polynomial of time, containing cross products of the time derivative of all of the quantities in the numerator and denominator of the $c_i(t)$. Without loss of accuracy, we retain only the linear terms in both the numerator and the denominator, yielding

\begin{equation}
[c_0(t)+c_l(t)(1+\tau^u(t))+c_q(t)(2(1+\tau^u(t))+\tau^u(t)^2)]e^{-\tau^u(t)}
\approx\frac{c_{e,1} t + c_{e,2}}{c_{e,3} t + c_{e,4}}e^{-\dot{\tau}^u t - \tau^u}
\label{eq:rational1}
\nonumber \\ 
\end{equation}
where the $c_i$ are extensive combinations of the average and time derivative of optical depth, opacity, and emissivity terms in nearby grid points. 

Similarly, we can reduce the final term in Equation~(\ref{eq:formalsolutionreorg}) to a simple rational form, 

\begin{equation}
c_0(t)+c_l(t) + 2 c_q(t)\approx\frac{c_{r,1} t + c_{r,2}}{c_{r,3} t + c_{r,4}}.
\label{eq:rational2}
\nonumber \\
\end{equation}
We note that Equations~(\ref{eq:rational1}) and (\ref{eq:rational2}) can be rewritten as expressions resulting from a Pad\'e approximation.
Now the function to evaluate becomes
\begin{equation}
I(t)\approx I_0(t)e^{-\tau^u(t)}-\frac{c_{e,1} t + c_{e,2}}{c_{e,3} t + c_{e,4}}e^{-\dot{\tau}^u t - \tau^u}+\frac{c_{r,1} t + c_{r,2}}{c_{r,3} t + c_{r,4}},
\label{eq:formalsolutionapprox}
\end{equation}
which is readily solved using
\begin{equation}
\int \frac{\xi-z_{e,1}}{\xi-p_{e,1}}e^{-\xi} d\,\xi =  
e^{-p_{e,1}} (p_{e,1} - z_{e,1}) {\rm {\bm Ei}}[p_{e,1} - \xi]-e^{-\xi},
\nonumber \\
\end{equation}
and
\begin{equation}
\int \frac{\xi-z_{r,1}}{\xi-p_{r,1}} d\,\xi=\xi + (p_{r,1} - z_{r,1})\ln(-p_{r,1} + \xi),
\nonumber \\
\end{equation}
to yield
\begin{equation}
\overline{I}(t)=\int_0^{\delta t}I_0(t)e^{-\tau^u(t)} d\,t+[ K_{e,0} Q_{e,0} \mid_0^{\xi_0} + K_{r,0} Q_{r,0}\mid_0^{\xi_1} ],
\nonumber \\
\end{equation}
with
\begin{equation}
Q_{e,0}= e^{-(\xi + \tau^u)} - (z_{e,1}-p_{e,1})(e^{-\tau^u-p_{e,1}}{\bm Ei}[\xi + p_{e,1})],
\nonumber \\
\end{equation}

\begin{equation}
Q_{r,0}=\xi  + (z_{r,1}-p_{r,1})
\,\,{\bm {\ln}} [\xi+p_{r,1}].
\nonumber \\
\end{equation}
Similarly, 

\begin{equation}
\hat{I}(t)=\int_0^{\delta t}t I_0(t)e^{-\tau^u(t)} d\,t +
[ K_{e,1} Q_{e,1} \mid_0^{\xi_0}+ K_{r,1} Q_{r,1} \mid_0^{\xi_1} ],
\end{equation}
where
\begin{eqnarray}
&& Q_{e,1}= - \frac{1}{\dot{\tau}_u} 
\times 
\left[(1-z_{e,1}+p_{e,1}) e^{-(\xi+\tau^u}) 
+ \xi\,\, e^{-(\xi+\tau^u)}-p_{e,1} (z_{e,1}-p_{e,1})(e^{\tau^u-p_{e,1}}{\bm Ei}[\xi + p_{e,1}) \right], 
\end{eqnarray}

\begin{eqnarray}
&&Q_{r,1} = K_{r,0} \left( \frac{1}{2} 
\xi(\xi+2 z_{r,1}-2 p_{r,1})
- p_{r,1} (z_{r,1}-p_{r,1}) \,\, {\bm{\ln}}[\xi+p_{r,1}] \right),
\end{eqnarray}
with 
\begin{eqnarray}
K_{e,0}=K_{e,1}=-\frac{c_{e,1}}{\dot{\tau}^u \,\,c_{e,3}},\\ \nonumber
K_{r,0}=K_{r,1}=\frac{c_{r,1}}{c_{r,3}},
\\ \nonumber
z_{e,1}=-\dot{\tau}^u\,\, \frac{c_{e,2}}{c_{e,1}},
\,\,\,\,\,\,\,\,\,\,
p_{e,1}=-\dot{\tau}^u\,\, \frac{c_{e,4}}{c_{e,3}},
\\ \nonumber
z_{r,1}=\frac{c_{r,2}}{c_{r,1}},
\,\,\,\,\,\,\,\,\,\,
p_{r,1}=\frac{c_{r,4}}{c_{r,3}},\nonumber \\
 \xi_0=\delta t\,\, \frac{c_{e,3}}{c_{e,4}}, \,\,\,\,\,\,
 \xi_1=\delta t.
\end{eqnarray}
This completes the calculation of the time-integrated formal solution.

\subsection{Propagation of the intensity}
All we need now to complete the calculation of $\overline{I}$ and $\hat{I}$ is to evaluate the first term on the right-hand side,
\begin{equation}
\int_0^{\delta t}I_0(t)\,e^{-\tau^u(t)} d\,t \,\,\,\,\,\,{\rm and}\,\,\,\,\,\, \int_0^{\delta t}t\,I_0(t)\,e^{-\tau^u(t)} d\,t.
\nonumber \\
\end{equation}
Unfortunately, $I_0(t)$ is the result of a spatial and frequency interpolation of the upwind intensity and does not have a simple form. In fact, it is straightforward to show that to retain the exact time dependence of the radiation field, for every propagation step across the grid, new terms in a sum of increasing length are introduced, quickly leading to an intractable expression.

To overcome this complexity, we approximate the intensity with a polynomial form, since this form will not grow in complexity when interpolated spatially. 
Since the weighting that this procedure will give over the interval of integration is not clear in general, we choose not to apply any weight  and to simply fit a polynomial of order $n$ with equal weights for all deviations from the true intensity across the time interval. In this simple case, the optimal coefficients are given by those that minimize the distance between the polynomial form and the true intensity over the time step.

We define the distance as the integral of the square of the difference between the two functions on the time interval $[0,\delta t]$,
\begin{equation}
\chi^2=\int_0^{\delta t} (I(t)-P_n(t))^2 dt,
\nonumber \\
\end{equation}
the minimum of which is located where the partial derivatives to the coefficients $c_n$ of the polynomial $P_n(t)$ vanish,
\begin{equation}
\frac{\partial}{\partial c_n} \chi^2=0.
\nonumber \\
\end{equation}
We substitute for $\chi^2$ and use the linearity of integration and differentiation,
\begin{equation}
\frac{\partial}{\partial c_n} \int_0^{\delta t} (I(t)-P_n(t))^2 dt =\int_0^{\delta t} \frac{\partial}{\partial c_n}(I(t)-P_n(t))^2 dt,
\nonumber \\
\end{equation}
to obtain
\begin{equation}
\int_0^{\delta t} -2\left[I(t)-P_n(t)\right] t^n dt=-2\left[\int_0^{\delta t} I(t) t^n dt - \int_0^{\delta t} P_n(t) t^n dt\right]=0.
\nonumber \\
\end{equation}
Interestingly, if we assume a linear form, $I(t)=P_1(t)=c_{0}\,t + c_{1}$,
and we minimize $\chi^2$, we obtain
\begin{equation}
c_{1} \frac{1}{3}\delta t^3 + c_{0} \frac{1}{2}\delta t^2 =\int_0^{\delta t}t I(t)dt \equiv \hat{I}(t),
\nonumber \\
\end{equation}
and
\begin{equation}
c_{1} \frac{1}{2}\delta t^2 + c_{0} \delta t=\int_0^{\delta t} I(t) dt \equiv \overline{I}(t),
\nonumber \\
\end{equation}
the right-hand side of which contains only moments of the time-integrated intensity, $\hat{I}(t)$ and $\overline{I}(t)$, both of which were already computed in all upstream grid points and can thus be obtained for free. In terms of these quantities, the coefficients of $P_1(t)$ are given by
\begin{equation}
c_{0}=\frac{2(2\overline{I}(t)\delta t-3\hat{I}(t))}{\delta t^2},
\nonumber \\
\end{equation}
\begin{equation}
c_{1}=\frac{6(2\hat{I}(t)-\overline{I}(t)\delta t)}{\delta t^3},
\nonumber \\
\end{equation}
which can now be used to spatially interpolate the intensity and calculate the last missing term.

\subsection{Solving the Nonlinear Rate System}
\label{non-linear-rates}
In this section, we describe the nonlinear rate system and its solution. For ease of discussion we first fix the values of temperature and electron density, and then we describe how do we include them in the nonlinear rate system.

As discussed in the previous sections for a given population density, the formal solution provides the radiation field quantities $\bar{I}$ and $\hat{I}$; by integrating these, we can obtain the radiative rates.  Using the fixed temperature and the electron density, we can obtain the collisional rates. For a given set of collisional and radiative rates, solving the system in Equation~(\ref{eq:ne-system}) provides the population density. Since the population densities and the radiation field are nonlinearly coupled to each other, we need to iterate between the formal solution and the nonlinear rate system in Equation~(\ref{eq:ne-system}). The most effective way to solve this nonlinear problem is by linearizing the rate system and using iterative methods. Here this linearization is done by evaluating the derivatives of $\overline{I}(t)$ and $\hat{I}(t)$ with 
respect to $\dot{n}_l$. In general, the derivatives of the radiation field to solve the nonlinear non-LTE problems are known as the approximate lambda operators, and the resulting iterative methods are known as approximate lambda iteration (ALI) methods \citep[see][and the references cited therein]{1973ApJ...185..621C,1973JQSRT..13..627C,1981ApJ...249..720S,1986JQSRT..35..431O,2003ASPC..288...17H}. Particularly for the multilevel case, various approaches have been described in the literature for obtaining the approximate lambda operators. The most important of them are the "complete linearization'' technique  \citep[see][]{1969ApJ...158..641A,1985JCoPh..59...56S,1991ASIC..341...39C} and the one that is described as the MALI formalism by \citet[][]{1991A&A...245..171R,
1992A&A...262..209R}. The former method leads to a slightly faster convergence; the latter method has recently found the most widespread use as it is relatively simple to implement.

Although direct evaluation of the derivatives appears simple enough, even if we only consider the dependence on local populations, the expressions become quite extensive. We therefore draw on the 
basic assumption of the MALI formalism and determine the derivative with respect to the local emissivity only. The resulting dependence describes the intensity variations adequately and produces a much simpler expression for the derivative than when all other dependencies are considered. For the acceleration, we follow the preconditioned MALI scheme. Here the main idea is to perform an operator splitting of the nonlinear radiation field term in the rate equation and to use use the local operator $\Psi^*$ for the acceleration \citep[][]{1991A&A...245..171R,1992A&A...262..209R}. 

For fixed values of temperature and the electron density, we can rewrite Equation~(\ref{eq:ne-system}) so that the system we actually solve has the form
\begin{equation}
{\bm A}^\star \cdot {\bm x} = {\bm b}^{\star},
\label{eq:ne-system-ndot}
\end{equation}
where the solution we look for is ${\bm x}=\{\dot{n}_1,\cdot\cdot\cdot,\dot{n}_N\}$, where $N$ is the total number of atomic energy levels considered. 
Following the preconditioned MALI scheme, we obtain the nonlinear terms of the form $\dot{n}_l \dot{n}_m$ in our system. They are linearized by using the "previous iteration solution,"  or "old" $\dot{n}^{\dagger}$, for one of the $\dot{n}$s and solve the linear system for the "new" $\dot{n}$. 
The choice of the particular $\dot{n}$ that is "old" or "new" determines the speed of the acceleration. 

When the temperature and electron densities are not fixed, the nonlinear system needs to be rewritten to include $\dot{n}_e$ and $\dot{T}$. The modified nonlinear system has the same form as Equation~(\ref{eq:ne-system-ndot}), but now we solve for ${\bm x}=\{\dot{n}_e,\dot{T},\dot{n}_1,\cdot\cdot\cdot,\dot{n}_N\}$. This general case leads to additional nonlinear products of the form $\dot{n}_l\,\dot{n}_e$, $\dot{n}_e \,\dot{n}_m$, $\dot{n}_l\,\dot{T}$ and $\dot{T} \,\dot{n}_m$. We use the same technique of linearizing these products by using the previous iteration solution, or old values, for one of the time derivatives in the product.

The form of the nonlinear rate system 
written in order to solve ${\bm n}$ as in Equation~(\ref{eq:ne-system}) (or 
rewritten in order to solve ${\bm x}$ as in Equation~(\ref{eq:ne-system-ndot}) )
is similar to the non-LTE multilevel kinetic equilibrium equation (see Equation~(\ref{eq:see})), but with a generalized form to take care of the time dependence. The important difference between the kinetic equilibrium equation and the nonequilibrium rate system is that that the rate matrix ${\bm A}$ and the vector ${\bm b}$ both nonlinearly depend on the solution ${\bm n}$ through the radiation field terms of the form $\bar{J}_{l,l^{\prime}}$ and $\bar{J}_{l^{\prime},l}$, with $l$ and $l^{\prime}$ being labels of energy level (see below).
A detailed description of the time-dependent MALI scheme with the acceleration is given in the Appendices \ref{appendixa} and \ref{appendixb}.
The final expression of the time-dependent rate system is given by
\begin{eqnarray}\label{rate-system-integrated-a} 
&& \dot{n}_l\,\delta\,t - 
\Big\{\sum_{l^\prime} \dot{n}_{l^\prime} n_e \hat{D}_{l^\prime\,l}-
\sum_{l^\prime} \dot{n}_l n_e \bar{D}_{l\,l^\prime}\Big\} \\ \nonumber 
&&-\Big\{\sum_{l^\prime}\dot{n}_{l^\prime}
\Big( \hat{U}_{l^\prime\,l}+ n_e \hat{U}^{\star}_{l^\prime\,l}
+\hat{J}_{l^\prime\,l}+ n_e \hat{J}^{\star}_{l^\prime\,l}\Big)\\ \nonumber
&&-\sum_{l^\prime} \dot{n}_l \Big(\hat{U}_{l\,l^\prime}+n_e \hat{U}^{\star}_{l\,l^\prime}
+\hat{J}_{l\,l^\prime}+n_e \hat{J}^{\star}_{l\,l^\prime} 
\Big)\Big\}\\ \nonumber
&&-\dot{n}_e\Big\{
\sum_{l^\prime} \Big(n_{l^\prime} \hat{D}_{l^\prime\,l}-n_l \hat{D}_{l\,l^\prime}\Big)
+\sum_{l^\prime} n_{l^\prime}\Big(\hat{U}^{\star}_{l^\prime\,l}+\hat{J}^{\star}_{l^\prime\,l} \Big) 
-n_l \Big( \hat{U}^{\star}_{l\,l^\prime}+\hat{J}^{\star}_{l\,l^\prime}\Big) 
\Big \}\\ \nonumber
&&-\dot{T} \Big \{
\sum_{l^\prime} \Big(n_{l^\prime} n_e \hat{d\,D}_{l^\prime\,l}-n_l n_e \hat{d\,D}_{l\,l^\prime}\Big) \\ \nonumber
&&+\sum_{l^\prime} n_{l^\prime}\Big(\hat{d\,U}_{l^\prime\,l}+n_e \hat{d\,U}^{\star}_{l^\prime\,l}
+\hat{d\,J}_{l^\prime\,l}+n_e \hat{d\,J}^{\star}_{l^\prime\,l} \Big) \\ \nonumber
&&-n_l \Big( \hat{d\,U}_{l\,l^\prime}+n_e \hat{d\,U}^{\star}_{l\,l^\prime}
+\hat{d\,J}_{l\,l^\prime}+n_e \hat{d\,J}^{\star}_{l\,l^\prime}\Big) 
\Big \}\\ \nonumber
&&+
\sum_{l^{\prime}} \dot{n}_{l^{\prime}} 
\Big(\hat{\Psi}^{\star}_{I,l^{\prime},l} 
+\hat{\Psi}^{\star}_{II,l^{\prime},l} 
+\hat{\Psi}^{\star}_{III,l^{\prime},l}
\Big)
-\Big(\sum_{l^{\prime}} \dot{n}_l \hat{\Psi}^{\star}_{Ia,l,l^{\prime}}
+\dot{n}_e \hat{\Psi}^{\star}_{IIa,l,l^{\prime}}
+\dot{T}\,\, \hat{\Psi}^{\star}_{IIIa,l,l^{\prime}} \Big)\\ \nonumber 
&&=\\ \nonumber 
&&\sum_{l^\prime}\Big(n_{l^\prime} n_e \bar{D}_{l^\prime\,l}-n_l n_e \bar{D}_{l\,l^\prime}\Big) \\ \nonumber
&&+\sum_{l^\prime} \Big(
n_{l^\prime} \bar{U}_{l^\prime\,l}+n_{l^\prime} n_e \bar{U}^{\star}_{l^\prime\,l}
+n_{l^\prime} \bar{J}_{l^\prime\,l}+n_{l^\prime} n_e \bar{J}^{\star}_{l^\prime\,l}
\Big)\\ \nonumber
&&-\Big( 
n_{l} \bar{U}_{l\,l^\prime}+n_{l} n_e \bar{U}^{\star}_{l\,l^\prime}
+ n_{l} \bar{J}_{l\,l^\prime}+n_{l} n_e \bar{J}^{\star}_{l\,l^\prime}
\Big) \\ \nonumber
&&+ \bar{\Psi}^{\star}_{IV,l} .
\end{eqnarray}
where $\dot{n}_{l,n-1}$ is the solution from the previous time step with time index $(n-1)$, and the various other terms are described in the Appendices (\ref{appendixa} and \ref{appendixb}).

In the steady-state case, the time evolution of the solution will settle toward the equilibrium solution. As the solution settles, the remaining difference between the current solution and the equilibrium solution decreases, and the time-step control gradually allows the time steps to become larger. Since the scattering rate is not dependent on the time step, as the time steps increase, more scatterings take place in each time step, resulting in a numerically ``stiffer'' problem. In the dynamic case, the time step is never allowed to become larger than the dynamic timescale; hence the number of scatterings remains roughly constant. As a result, in the dynamic case, the system is expected to be less stiff.

The solution $\dot{n}_l$ of the rate system in Equation~(\ref{rate-system-integrated-a}) is used to update the opacity and emissivity, from which the updated $\bar{I}$ and $\hat{I}$ are then computed. Integrating them, the rates can be updated, and Equation~(\ref{rate-system-integrated-a}) can be solved. These two steps are iterated until convergence is obtained. In this way, the time dependence in our formalism is implicit, as the nonlinear timedependence appears through iteration between the formal solver and the rate system. 

When the temperature is not fixed, it also becomes part of the solution of the non-linear rate system (see above).
To this end we combine the atomic rate system with the molecular rate equations (see Section \ref{molecules} below), the energy balance equation (see Section \ref{eos}) and finally the Saha-Boltzmann equations to treat the metal equations.
The nonequilibrium treatment in the rate system is applied always to the hydrogen atom and the hydrogen molecules. For all the other elements, a provision is made for them to be treated either in LTE, or in non-equilibrium similar to the treatment of the hydrogen atom itself. 

\subsection{Nonequilibrium Molecular Rates}
\label{molecules}
Molecular formation/destruction takes place through multiple paths, which together are known as a chemical reaction network. This reaction network for each molecule results in an additional equation in our rate system that has the general form
\begin{eqnarray}
\label{eq:mol-neq}
&& \frac{\partial {N_{{\rm mol,i}}}}{\partial t}+({\bm v}\cdot\nabla){N_{{\rm mol,i}}} = \sum_{l} {N_{{\rm mol,i}}} {N_{{\rm mol,j}}} K^+_l
-{N_{{\rm mol,{i^{\prime}}} }} {N_{{\rm mol,{j^{\prime}}}}} K^-_l,
\end{eqnarray}
where $K^+_l$ and $K^-_l$ are the reaction rates for the $l$th chemical reaction, and $N_{{\rm mol,i}}$,$N_{{\rm mol,j}}$,$N_{{\rm mol,{i^{\prime}}}}$ and $N_{{\rm mol,{j^{\prime}}}}$ are the reactants.
We have also treated molecular formation and dissociation of 
${\rm{H_2}}$, ${\rm{H_2^+}}$ and ${\rm{H^-}}$ also in nonequilibrium by considering several collisional
and radiative chemical reactions and their rates. However, in this paper, we
restrict ourselves to stationary solutions for fixed molecular densities, metal densities, and temperature. A detailed study of the role of nonequilibrium treatment of these molecules is deferred to a subsequent paper.

\subsection{Coupling to Hydrodynamics}
The coupling of the RT and the hydrodynamics is through the mass density $\rho$, velocity $v$, gas pressure $P$ and the radiative flux divergence $F$. At the beginning of each step, the mass density $\rho$ per time step is converted into number densities of the total hydrogen in atomic and molecular form and the number densities of other elements assuming their abundance to be fixed. This total hydrogen in all forms is used for the particle conservation equation. From the solution of the RT equation, we have $\bar{I}$, which can be readily used to calculate the radiation flux $F$. From the solution of the rate system, we have $\dot{n}_l$, $\dot{n}_e$ and $\dot{T}$; using these, the population $n_l$, the electron density $n_e$, and the temperature $T$ per time step are updated, and their partial pressures are computed, the total of which forms the total gas pressure $P$. The velocity $v$ per time step is used for advecting the $\dot{n}_l$ and for the Doppler shifts of the spectral lines.
For the MHD part, we use the original Max Planck Institute for Solar System Research, University of Chicago Radiation MHD (MURaM) solver.
In the dynamic case, for each time step, the RT and the MHD solvers are iterated until converged. This iteration is the outermost iteration loop in addition to those involving RT quantities.
Since all of these are not relevant for the stationary case, further details and testing will be provided in a forthcoming paper dedicated to the dynamic case.

\begin{figure*}[ht]
\includegraphics[scale=0.5]{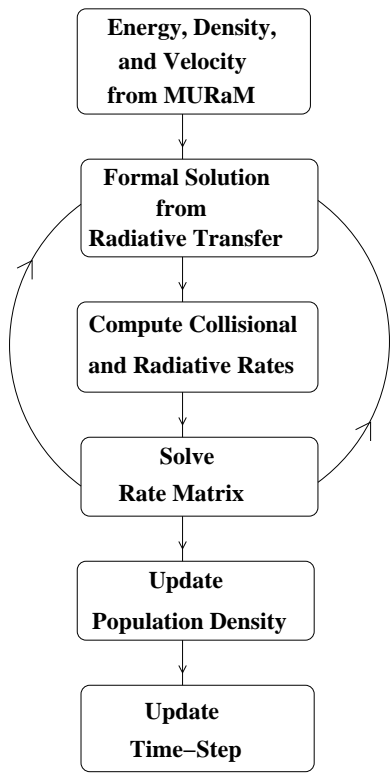}
\caption{Flow diagram showing the RT iterative scheme for a given time step.}
\label{fig:flowchart}
\end{figure*}
\section{Implementation}
\label{sec:implementaion}
The scheme described above is implemented as a module in the MHD simulation code MURaM. The code is written in C++. The module is written in 1D/2D/3D but currently tested only in 1D. 
The results from the future extension of our work to 2D/3D will be presented in subsequent papers.
We stress here that for the current paper, we restrict ourselves to a study of steady-state solutions with
the aim of verifying that our RT module works and provides correct results.
The RT module iteratively solves
the formal solution of the RT equation using a short-characteristics formal solver with a choice of 1D/2D/3D Cartesian geometry, self-consistently and iteratively with the 
time-generalized Rybicki-Hummer MALI scheme (see Section~\ref{non-linear-rates}, Appendices~\ref{appendixa} and \ref{appendixb})
that solves the rate system, for a multilevel atom system. The formal solution involves evaluation of the local and propagation parts of the of the $\bar{I}$ and $\hat{I}$ in different optical depth regimes.
The RT solver is also implemented to handle velocity fields.
A flow diagram of the important steps of our RT scheme is presented in Figure~\ref{fig:flowchart}.

\subsection{Numerical Details}
An important step in the implementation is that we approximate the time integrals of the time-dependent
intensity to obtain $\bar{I}$ and $\hat{I}$. As already described in 
Section~\ref{sec:theory}, this is analogous to the  
integration of the spatially dependent source function in the original short-characteristics formal solution method.
The integrated expressions thus obtained
contain exponential integrals that need to be evaluated numerically. Further, 
for different optical depth ($\tau$) values (e.g., for small $\tau$), 
we expand the polynomial form of $I(t)$ using various series 
expansions to avoid numerical cancellations. In these cases, the exponential integrals simplify
to become polynomial expressions. \\

\begin{figure*}
\includegraphics[scale=0.5]{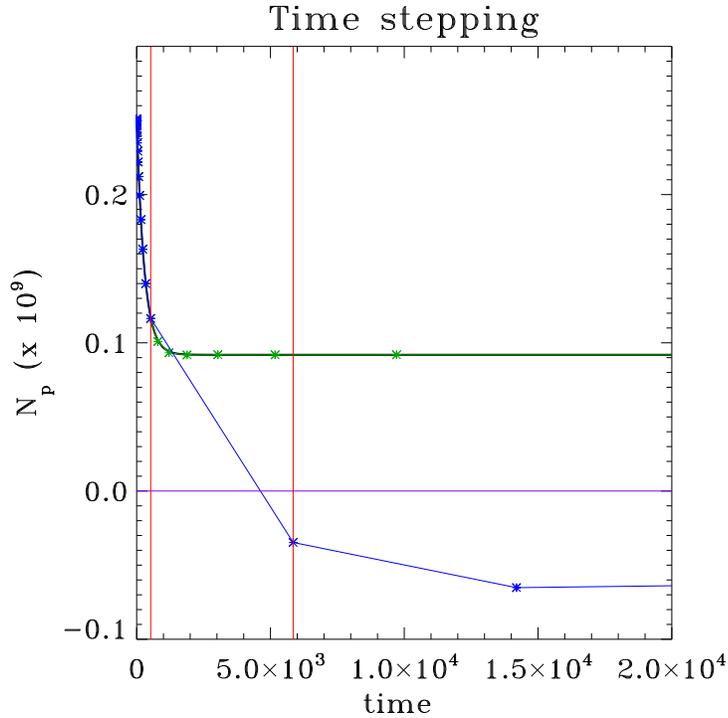}
\caption{Illustrative time evolution of the proton density $N_p$ at a fixed column mass using very fine (black line; $f=0.01$) and intermediate (green line and stars; $f=0.05$) time-stepping. These two curves are visually indistinguishable  as the difference between them is very small. An example of a very large $\delta t$ leading to negative proton density is also shown (blue line and stars). The two vertical lines are drawn to indicate the initial and final time values of this large $\delta t$ (red lines). The zero line is also drawn for reference (purple line). For further details, see Section~\ref{sec:time-step}.
}
\label{fig:time-stepping}
\end{figure*}

\subsection{Time-step Criterion}
\label{sec:time-step}
In this section, we discuss the time-stepping criterion we adapted in our RT scheme. Since we solve the rate system for $\dot{n}$, and we assume a linear time dependence for the populations, we impose a criterion of the form
\begin{equation}
{\delta\,t}_{i+1} = f \times \min\{(n_i+{\delta\,t}_i \dot{n}_i)/\dot{n}_i\},
\label{eq:time-stepping}
\end{equation}
where $i$ and $i+1$ are the indices of two successive time steps, and a minimum is taken over all of the levels and the spatial dimensions. 
Here $f$ is a fraction taken to be a free parameter less than unity that determines how crude or fine the time resolution is. For example, $f=0.01$ means that the change in the population is 1\% of the population itself. 
Since we solve for $\dot{n}$ at each time step, we do not know the value of $\dot{n}_{i+1}$ at the beginning of the $(i+1)$th time step. Therefore, $\dot{n}_i$ itself is used for estimating $(\delta\,t)_{i+1}$. Further, to ensure a smooth evolution, we take a linear combination of two successive $\delta\,t$ estimates (steps $i-1$ and $i$) as the actual $\delta\,t$ for the $i$th step. The time stepping so defined ensures that the change ${\delta n}_i=\delta\,t_i \dot{n}_i$ is small enough compared to $n_i$, which is important for a stable evolution of the system.

At every time step, the value of $\delta t$ is set by the largest $\dot{n}$ over all the levels and all the spatial points. In other words, $\delta t$ is set by the largest change in the system. Once the population $n$ at a given level and a given spatial point reaches its equilibrium value, the corresponding $\dot{n} \rightarrow 0$, and consequently, at this level and spatial point,  $\delta t \rightarrow \infty$. However, the criterion searches for the smallest $\delta t$ which is now set by the new largest $\dot{n}$ that corresponds to a different level and/or a different spatial point. In this way, $\delta t$ automatically takes larger and larger values as the $\dot{n}$ at different levels and spatial points reaches its equilibrium values.

In Figure~\ref{fig:time-stepping} we show the time evolution of the proton density denoted by $N_p$ at a fixed column mass for various time-stepping criteria. To represent very fine and intermediate time resolutions, we show cases with $f=0.01$ and $0.05$ (see Equation~\ref{eq:time-stepping}). These values respectively correspond to the change in population amounting to 1\% and 5\% of the $N_p$. Since the differences between the $N_p$ curves for these two cases are small, they are visually indistinguishable. We also show an example of $N_p$ evolution that has $f=0.05$ up to a point in time of $\sim$ 790 s, after which it is extrapolated using a large $\delta t$. This leads to negative $N_p$ values in the next time step. This shows the importance of sufficiently small time-stepping criteria, which otherwise can lead to unphysical, negative populations. Therefore, in order to make $\delta t$ sufficiently small, it is necessary to define the time-stepping criterion that depends on $\dot{n}$ at the end of every time step, as in Equation~(\ref{eq:time-stepping}).

\subsection{Computational demands}
We stress here that the purpose of this paper is to introduce our formalism and to show that it provides the correct results; consequently, the code has not yet been extensively optimized. As discussed below, the computational demands of the newly developed nonoptimized RT module are much higher when compared to that of the original MURaM. The increase in the computational costs has two main causes : (1) the increased wavelength dependence of the radiation field and the related quantities and (2) the requirement of the use of the long double data type for a precise evaluation of the exponential integrals and solving the resulting rate system. While the former cannot really be optimized extensively, the latter can possibly be dealt with by means of a more appropriate formulation of the relevant expressions.

For the convenience of the discussion, we define CPU time for convergence as the computing time required to complete the convergence cycle and
reach a fixed level of accuracy. For this purpose, we define the maximum relative error as 
\begin{equation} \label{eq:rel_err}
R_c=\max_{i,k}{\left | \frac{(n^{(m,i,k)}-n^{(m-1,i,k)})}{n^{(m,i,k)}}\right |},\,\,\,\,\,\, \\ \nonumber 
\end{equation}
where $n^{(m,i,k)}$ is the population in the $m$th iteration, for the $i$th energy level and $k$th grid point in a given time step. We then impose at each time step the maximum relative error $R_c$ to be less than a given parameter $\epsilon$, which is generally known as the convergence criterion. We used $\epsilon=10^{-4}$ for the results presented in this paper.

For example, for a 1D atmosphere with an extent of 2.8 Mm, represented using 151 spatial points, using a three-level hydrogen atom with a total of 90 wavelength points (including the line and the continuum), the CPU time required by the nonoptimized code is 2 s per iteration per time step on a single core of an Intel Xeon Gold 6150 CPU running at 2.70GHz. The total time per time step depends on the total number of iterations needed per time step, which is larger when the $\delta t$ is larger. Here we chose a fine time resolution of $f=0.01$. Typically, in the above example, when $\delta t\sim 10^{-3}$, a total of 30-40 iterations per time step are needed to reach convergence, resulting in an execution time of about 2 minutes per time step.

\section{Verification}
\label{sec:verification}
In this paper, we focus on the stationary solutions.
We have thoroughly tested our code for the accuracy of the method and validated it by reproducing the benchmark results from earlier papers. 
For all of the results presented in this paper, we use the average BIFROST \citep[][]{2011A&A...531A.154G} atmosphere from one of the publicly available snapshots and keep the density and energy fixed, but we evolve populations. For a fixed mass density and energy at every depth point, we solve the LTE chemical equilibrium equation and also obtain a temperature and pressure consistent with this energy and mass density (solving for the initial equation of state). Other than atomic hydrogen, ${\rm{H}}$, we have hydrogen molecules ${\rm{H_2}}$, ${\rm{H_2^+}}$, ${\rm{H^-}}$ and several other elements while solving for the initial LTE equation of state.
The initial energy, mass density, velocity from the MHD part of the MURaM, temperature, molecular densities, and the metal densities from the initial LTE equation-of-state solution are kept fixed for all the studies in this paper. A study of the molecular rates and dynamic solution is reserved for separate forthcoming papers.

In Section~\ref{sec:benchmark} we show that our nonequilibrium solver evolves the populations to an equilibrium solution,
which closely resembles the non-LTE kinetic equilibrium solution computed by the well-tested RH code \citep[][]{2001ApJ...557..389U}.
In Section~\ref{sec:accuracy}, the accuracy of the linear approximation of the time dependence of the populations is explored. Finally, in Section~\ref{sec:ir-timescales}, we reproduce the time-scales from previous studies that used a different numerical method \citep[][]{2002ApJ...572..626C}.

\begin{figure*}[ht]
\includegraphics[scale=0.8]{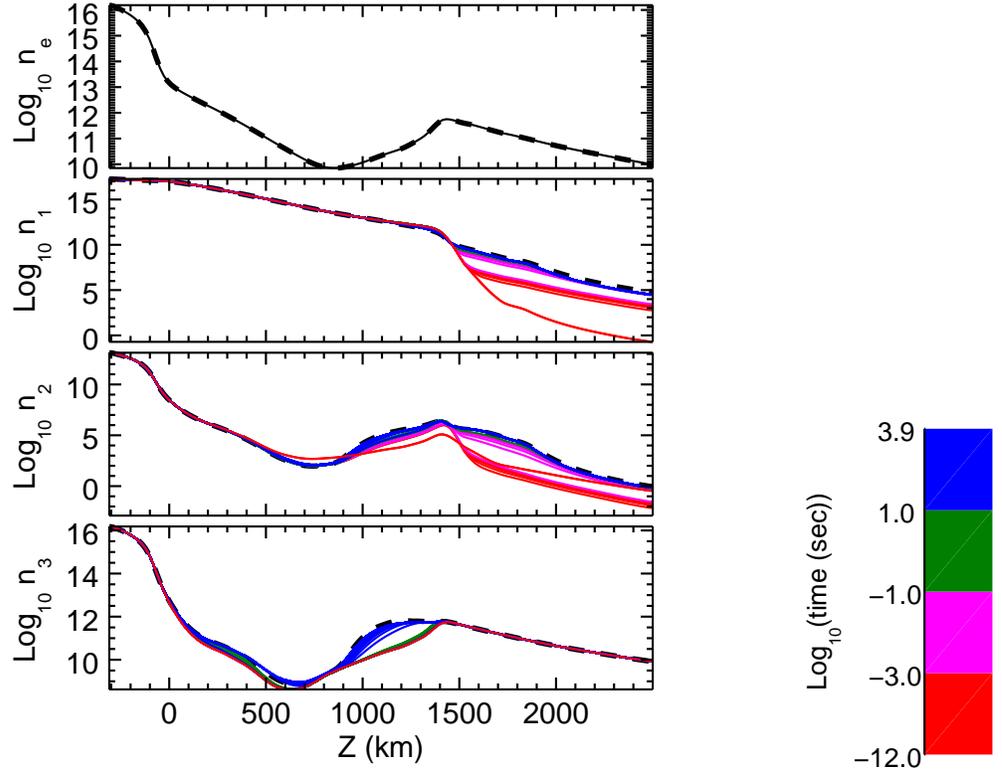}
\caption{Comparison of the equilibrium solution of the population evolution for the stationary case (solid lines) with the kinetic equilibrium solution from the RH code  
\citep[dashed lines; see][]{2001ApJ...557..389U}. The panels correspond to electron density $n_e$ (fixed) and three energy levels of hydrogen ($n_i$).  Different colors represent different time domains: red: $10^{-12}-10^{-3}$ s, magenta: $10^{-3}-0.1$ s, green: $0.1-10$ s and
blue: $10-10^{3.9}$ s. Selected curves representing each time interval are plotted.}
\label{fig:benchmark1}
\end{figure*}

\subsection{Benchmark Test}
\label{sec:benchmark}
In Figure~\ref{fig:benchmark1} we show the evolution of populations in a three-level ${\rm{H}}$ atom setup. We initialize the evolution of populations with their LTE values. When they eventually reach equilibrium, their values correspond to the instantaneous kinetic equilibrium solution of a non-LTE RT problem. To show that we indeed get a correct equilibrium solution, we overplot the solution from the non-LTE spectral synthesis code RH \citep[][]{2001ApJ...557..389U}, with the same atmospheric structure, atomic data, and background opacity. In addition to the fixed atmospheric structure, in both codes, we fix the electron density computed from an initial LTE equation-of-state solver. We obtain a very good match between the RH code solution and ours for all the bound (ground and first excited) and free (ionized) states. 

\begin{figure*}
\includegraphics[scale=0.4]{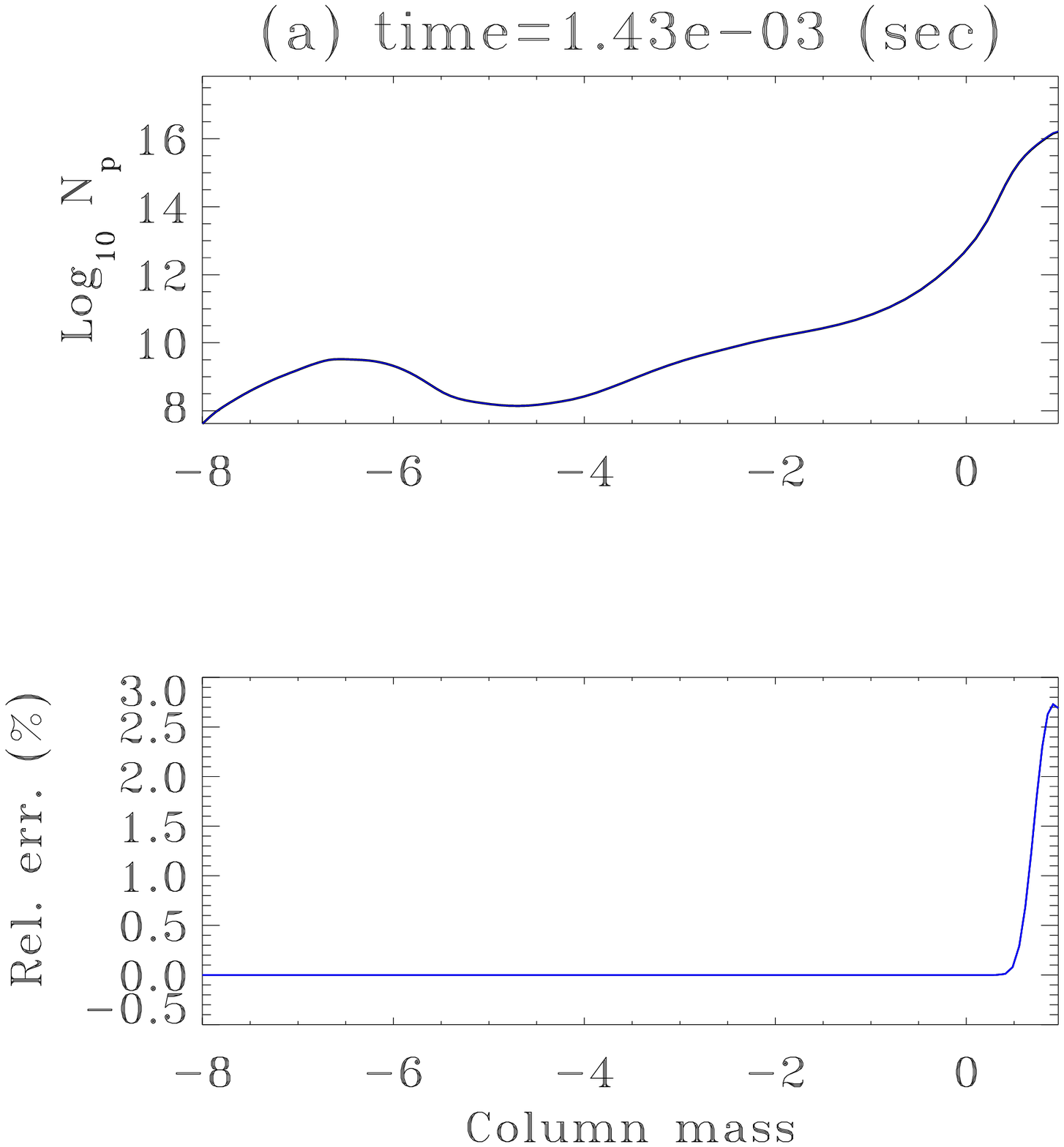}
\includegraphics[scale=0.4]{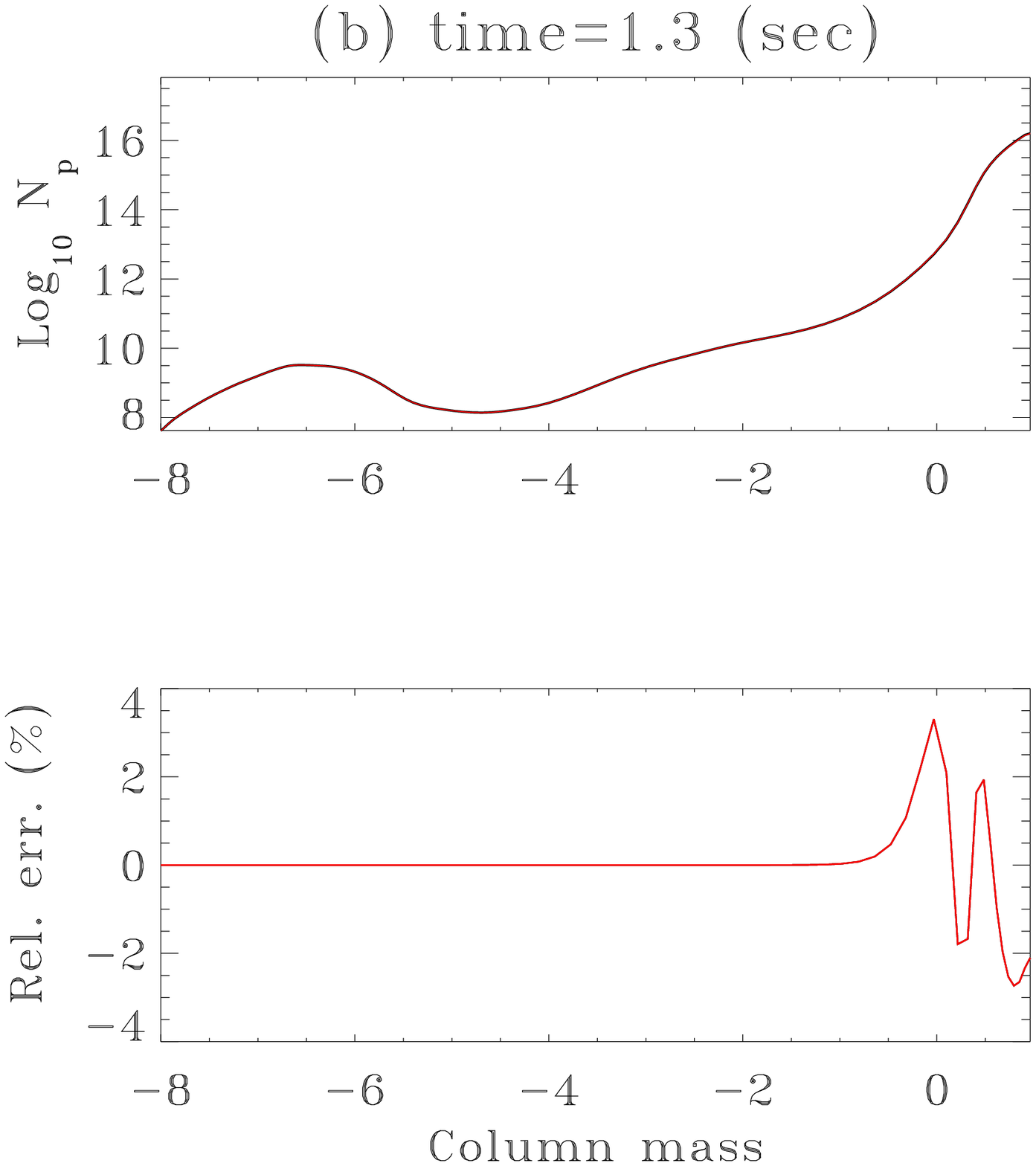}
\caption{Effect of crude time stepping on the population as a function of column mass for two chosen test-cases, shown as panels (a) and (b). 
Top: proton number density $N_p$ as a function of column mass for a fine time-stepping criterion of $f=0.01$, shown as black lines, in comparison to those with a crude time-stepping criterion of $f=0.5$ ($f=0.8$), shown as blue (red) lines. The curves are visually indistinguishable, as the difference between them is very small relative to scale on the axes.
Bottom: error in proton number density for the crude time stepping 
relative to that of the fine time stepping (described for the top panels) in percent as a function of column mass. All the plots have been shown for a fixed time at which the relative error reaches a maximum value (see title labels). 
}
\label{fig:time-step-crude1}
\end{figure*}
\begin{figure*}
\includegraphics[scale=0.4]{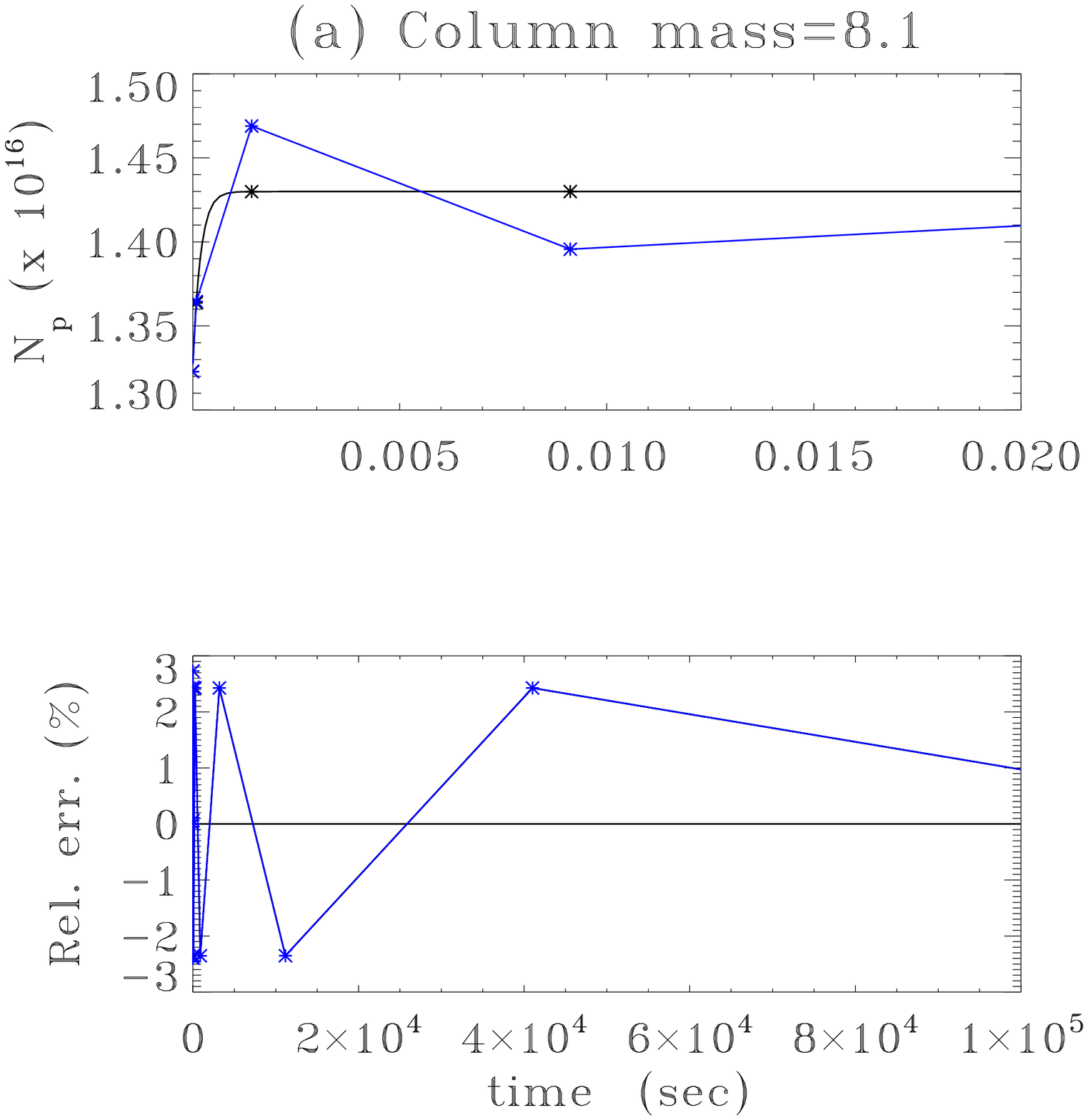}
\includegraphics[scale=0.4]{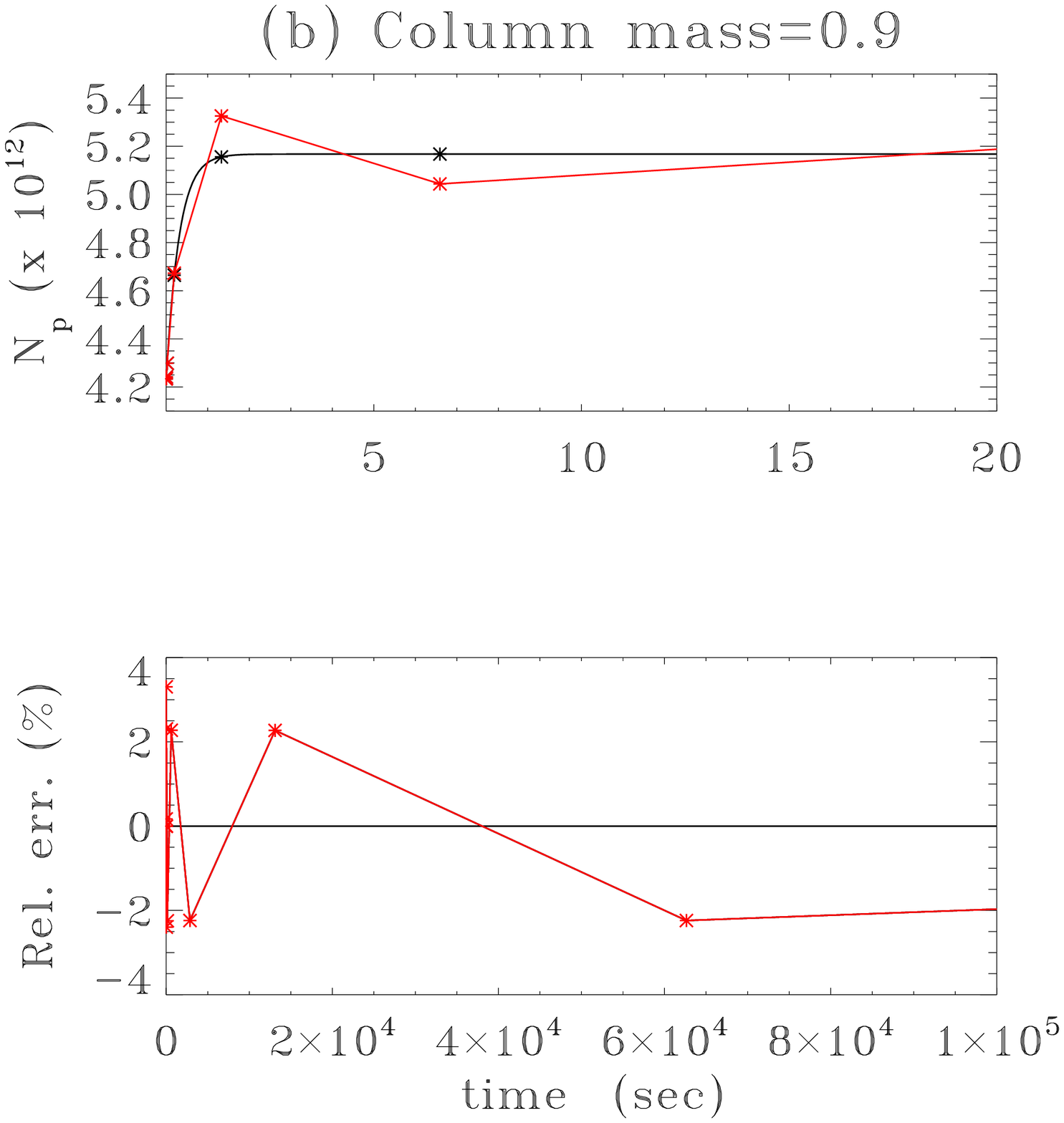}
\caption{Effect of crude time stepping on the population as a function of time for two chosen test cases, shown as panels (a) and (b). 
Top: proton number density $N_p$ as a function of time for a fine time stepping criterion of $f=0.01$, shown as black lines, in comparison to those with a crude time stepping criterion of $f=0.5$ ($f=0.8$), shown as blue (red) lines.
Bottom: error in $N_p$ for the crude time stepping 
relative to that of the fine time stepping (described for the top panels) in percent as a function of time.
All the plots have been shown for a fixed column mass at which the relative error reaches a maximum value (see title labels). 
}
\label{fig:time-step-crude2}
\end{figure*}

\begin{figure*}
\includegraphics[scale=0.4]{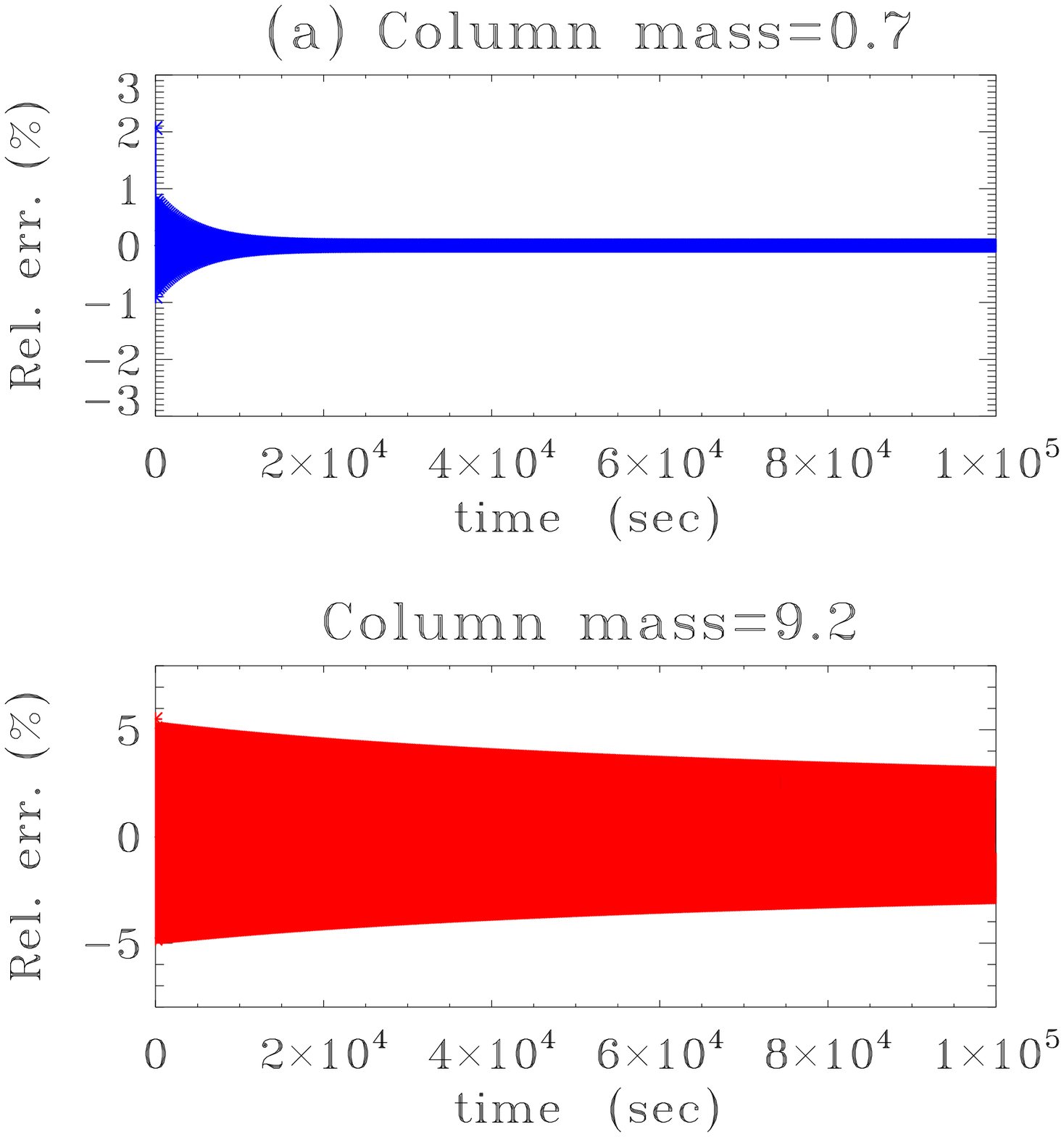}
\includegraphics[scale=0.4]{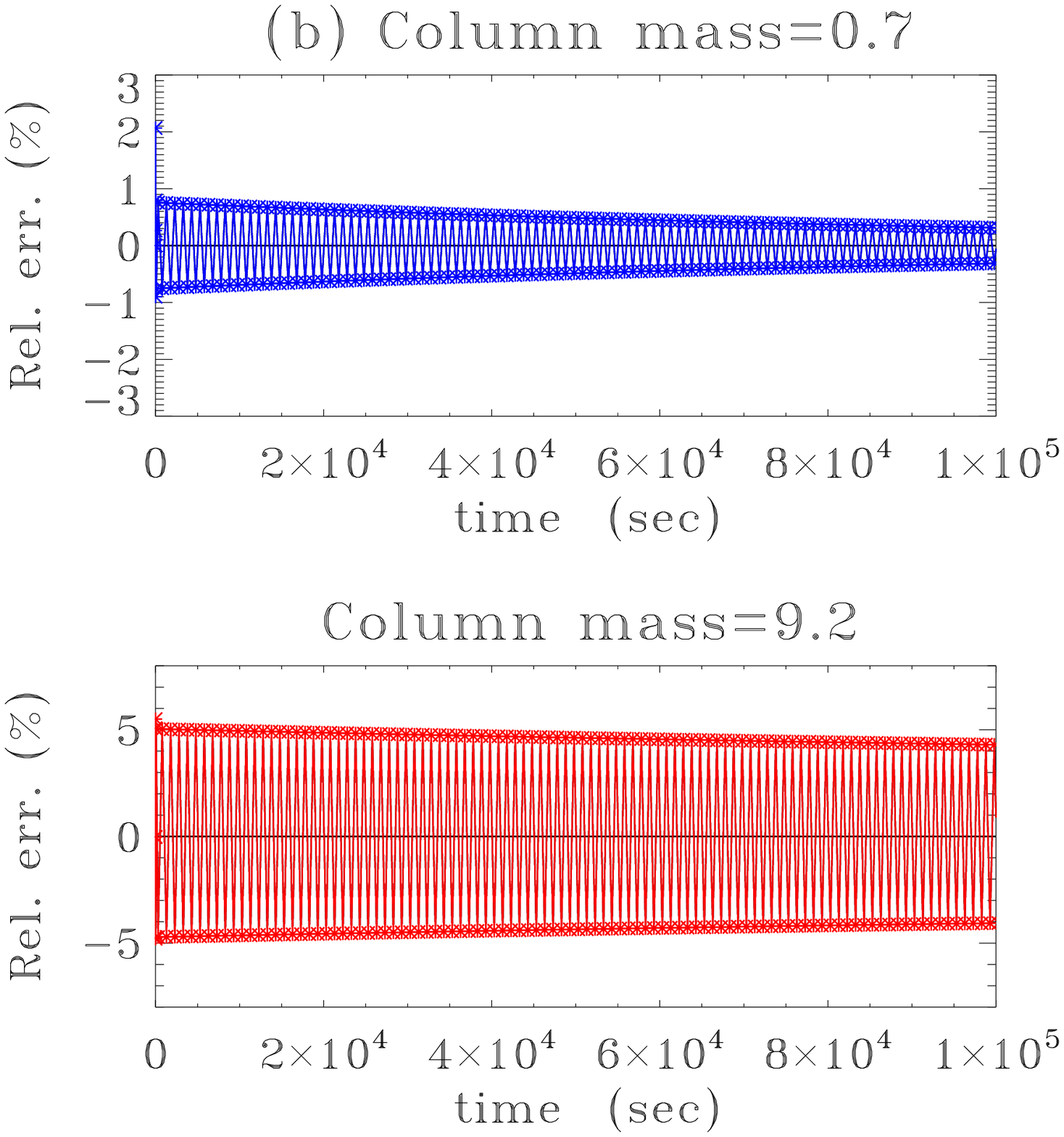}
\caption{Effect of crude time stepping on the population as a function of time for two chosen test cases, with both the cases having an upper limit on the $\delta t$.  They are $\delta t_{\rm max}=100$ (panel (a)) and $\delta t_{\rm max}=500$ (panel (b)). 
Top: error in proton number density $N_p$ for the crude time stepping ($f=0.5$)
relative to that of the fine time stepping ($f=0.01$) in percent as a function of time (blue lines).
Bottom: error in $N_p$ for the crude time stepping ($f=0.8$)
relative to that of the fine time stepping ($f=0.01$) in percent as a function of time (red lines).
All plots have been shown for a fixed column mass at which the relative error reaches a maximum value (see title labels). The upper limit of $\delta t$ is taken for both crude and fine cases.
}
\label{fig:time-step-crude3}
\end{figure*}

\subsection{Accuracy of the Linear Approximation}
\label{sec:accuracy}
Left to themselves, atomic populations away from the equilibrium state evolve to reach a steady state (or equilibrium state), which can be mathematically shown to follow an exponential law as a function of time, namely,
\begin{equation}
n(t) \sim a_0 \exp({-a_1\,t})+a_2,\\ \nonumber
\end{equation}
where $a_0$, $a_1$ and $a_2$ are constants. Using the above form in RT equation is not practical to solve. Instead, as described in Section~\ref{sec:popevolution} we use a piecewise linear 
function to describe the time dependence of the populations. Therefore accuracy of this description depends on the time resolution in our numerical setup (in other words, the value of $f$).

Now we analyze the accuracy of the linear time dependence of the populations.  We use a two-level atom model for this study, with only ground and ionized states. We compare the $N_p$ that used time resolutions that were very fine ($f=0.01$ in Equation~(\ref{eq:time-stepping})), and very crude ($f=0.5$ and $0.8$ in Equation~(\ref{eq:time-stepping})). In other words, $f=0.01$ corresponds to $\delta\,N_p=(\delta\,t)\, \dot{N}_p$; namely, the change in $N_p$ is 1\% of $N_p$, and $f=0.5$ ($f=0.8$) corresponds to the case where $\delta\,N_p=\delta\,t \dot{N}_p$ is 50\% (80\%) of $N_p$. 

These $N_p$ are shown as a function of column mass in Figure~\ref{fig:time-step-crude1} (top panels). In the bottom panels, we show the relative errors between the two $N_p$ curves in the top panels. All the plots correspond to a fixed time where the relative error is maximum. The maximum relative error is $\sim 3$ \% in the left panels and it is $\sim 4$ \% in the right panels. Thus, even with such extremely crude time resolutions the method remains stable producing relative errors of only a few percent. 

In Figure~\ref{fig:time-step-crude2} (top panels), 
we plot $N_p$ for the same fine and a crude time-stepping criteria as in Figure~\ref{fig:time-step-crude1} but as a function of time. In the bottom panels, we plot the relative error in percent between the two $N_p$ curves shown in the top panels. All the plots correspond to a fixed column mass where the relative error is maximum. We observe here that, once $N_p$ reaches its equilibrium value, $N_p$ computed using the crude time-stepping criterion oscillates around the value of $N_p$ computed using the fine time-stepping criterion at alternate time steps. As a consequence, as seen in the bottom panels, the relative error also oscillates, which never settles to zero. This is in contrast to a general expectation for the amplitude of the oscillations and the corresponding error to become smaller and smaller with time. We attribute this behavior of the error to the implicit interdependence of the solution $\dot{n}$ and the time step $\delta t$ as defined by the time-stepping criteria (Equation~(\ref{eq:time-stepping})). 

To understand this behavior of the error, in Figure~\ref{fig:time-step-crude3}, we plot the evolution of the relative errors between fine ($f=0.01$) and crude time-stepping ($f=0.5$ and $0.8$ for the top and bottom panels, respectively) time stepping with both cases having an upper limit on the time step of $\delta t_{\rm max}=100$ (panel (a)) and $\delta t_{\rm max}=500$ (panel (b)). We observe here that when the time step is fixed to a given value, the relative error continues to oscillate but with an amplitude that gradually decreases with time, and it saturates to a smaller value. This saturation value of the error is smallest for $f=0.5$, $\delta t_{\rm max}=100$, which corresponds to the finest time stepping among all four crude time-stepping criteria shown in this figure. Thus, the saturation value of the error determines the accuracy of the time stepping. 

Thus, we show here that the assumption of the linear approximation of the time dependence of populations in our method produces solutions that are accurate and robust against changes in the time resolution.

\begin{figure*}
\includegraphics[scale=0.7]{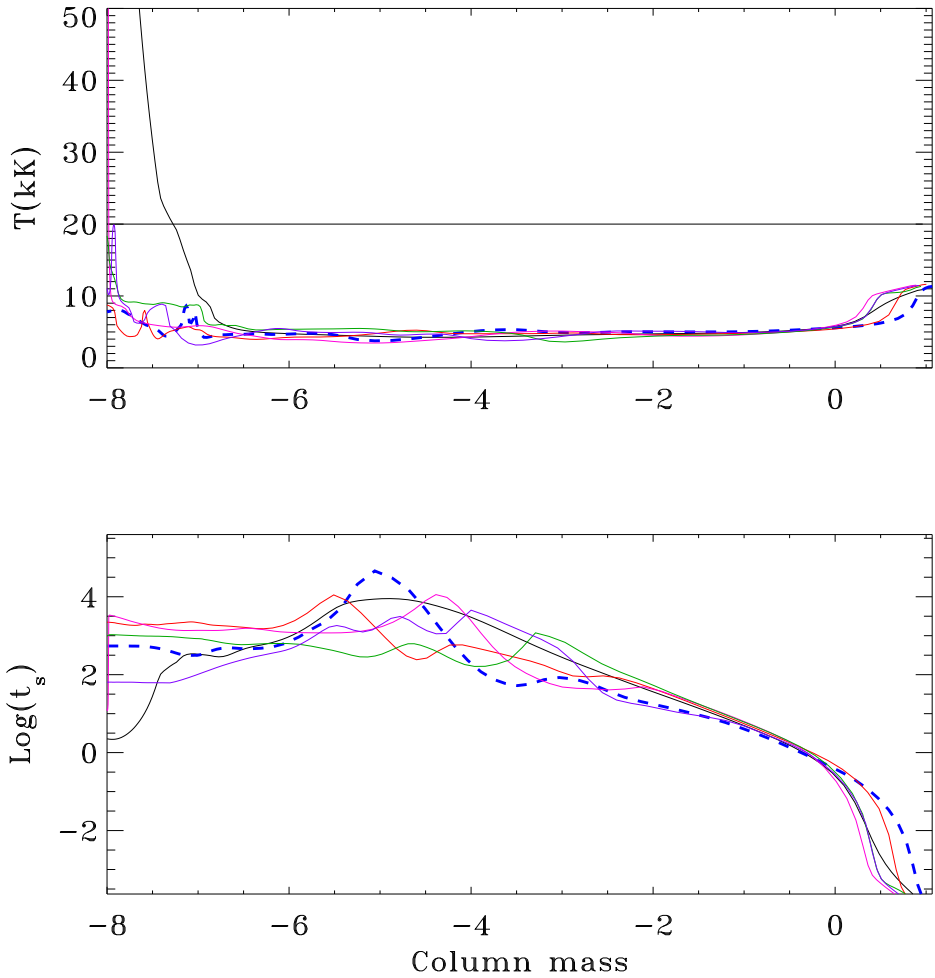}
\caption{Temperature stratification (top) and equilibrium timescale $t_s$ (bottom) for a number of different atmospheric structures. The case with the maximum timescale and corresponding temperature structure is shown as a blue dashed line.}
\label{fig:benchmark3}
\end{figure*}

\subsection{Ionization/recombination time scales}
\label{sec:ir-timescales}
In Figure~\ref{fig:benchmark3} we show different temperature structures and the corresponding timescale of ionization/recombination as a function of column mass. We follow the same approach as \citet[][]{2002ApJ...572..626C}. We consider a two-level atom model with only a ground state and an ionized state. We first start with an LTE solution and let our nonequilibrium solver reach equilibrium for a fixed atmospheric structure that defines an initial equilibrium proton density $N_p(0)$. We perturb the temperature by increasing its value by 1\% throughout the atmosphere. We then let the populations settle
to this new temperature structure. This equilibrium proton density is denoted as $N_p(\infty)$. The time evolution of the proton density from the initial state to the final state is represented by $N_p(t)$. We recall that the physical relaxation timescale, denoted here as $t_s$, of any system is defined as the
time it takes to change by a value of $e$. In particular, for computing relaxation timescale of $N_p$, we perform a least-squares linear fit to the function 
\begin{equation}
    \label{eq:time-scale}
\frac{N_p(t)-N_p(\infty)}{N_p(0)-N_p(\infty)}=
\exp(q_0+q_1\,t),
\end{equation}
where $q_0$ and $q_1$ are coefficients of the least-squares fitting with the relaxation timescale defined as $t_s=-1/q_1$ \citep[see also][]{2002ApJ...572..626C}.

We find that for various atmospheric structures that we have
chosen (density, energy, and corresponding temperature), the maximum time-scales can vary by orders of 
magnitude, from $\sim 100$ to $\sim 10^{4.5}$ in the mid-chromosphere and higher. This dependence of the time-scales on the atmospheric structure is caused by the strong density dependence of the collision rates that determine the timescales \citep[see][for a detailed discussion]{2002ApJ...572..626C}. 
The maximum time-scale that we obtain is $\sim 10^{4.5}$ (shown as dashed blue lines in Figure~\ref{fig:benchmark3}), which is quite close to the long timescales of $\sim 10^{5}$ obtained by \citet[][Figure 6]{2002ApJ...572..626C}. However, the timescales 
are small when we are in the photosphere, due to equilibrium conditions. 

\section{Summary}
\label{sec:summary}
In this paper we focus our attention on developing a numerical method to solve the non-LTE nonequilibrium RT problem through 
(a) a proper time-dependent treatment of the radiation
field,
(b) a proper nonequilibrium treatment of the molecular
chemistry, and
(c) the development of a time-implicit numerical scheme. The method is based on an integral equation approach to the RT equation that involves a generalization to the time dimension of (i) the short-characteristic technique for the formal solution of the RT equation and the (ii) MALI technique to solve the nonlinear rate system.

We validate our newly developed method with two important benchmark tests: (i) we start with LTE populations on a fixed atmospheric structure, allow them to evolve to the equilibrium solution, and verify that this agrees with the kinetic equilibrium solution obtained from the RH code \citep[][]{2001ApJ...557..389U}; and (ii) we show that the physical timescales required to reach equilibrium are similar to those obtained by \citet[][]{2002ApJ...572..626C}, who used a different numerical method. We also show that the solver remains stable and the solution is robust against changes in the time resolution.

The final aim of this work is to integrate this module with the radiation MHD code MURaM to carry out fully dynamic evolution of the MHD quantities and the RT quantities. This work is in progress, and the studies of the dynamic solution will be presented in a subsequent paper.

\acknowledgements{L.S.A. would like to thank Prof. Sami K Solanki and the European Research Council (ERC) under the European Union’s Horizon 2020 research and innovation program (grant agreement No. 695075) for funding this project.
}

\appendix
Here we present some important equations for solving the time-dependent rate system.
For simplicity of notation, we omit the dependence of all of the quantities
on $\bm{\Omega}$ and $\nu$ in general, and write it only when necessary. We start by presenting the rate equations in appendix \ref{appendixa}, the solution of which must generally be obtained by means of an iterative solver, after which, in appendix \ref{appendixb} we derive the expressions needed for the acceleration of the convergence of the iterative solver by means of operator splitting, a method commonly referred to as ALI \citep[see][and the references cited therein]{1973ApJ...185..621C,1973JQSRT..13..627C,1981ApJ...249..720S,1986JQSRT..35..431O,2003ASPC..288...17H}. Specifically, we implement here a variation on the MALI scheme \citep[][]{1991A&A...245..171R,1992A&A...262..209R}. For a given time step, we iteratively solve for $\dot{n}$, $\dot{n}_e$, and $\dot{T}$ where we note that the nonlinear dependence of the radiation field $I(t)$ on the temperature is treated implicitly, by using ``old'' temperature derivatives $\dot{T}^{\dagger}$ (which are solved for, iteratively, in the previous iteration) while solving for the current $\dot{n}$.

\section{Time-dependent Rate Equations}
\label{appendixa}
We can write the time-dependent rate system as
\noindent
\begin{eqnarray}
&&\sum_{l^\prime}n_{l^\prime}(t) n_e(t) C_{l^\prime\, l}(T(t)) \\ \nonumber
&& +\sum_{l^\prime}\oint d\Omega \frac{d\nu}{h\nu}[n_{l^\prime}(t)U_{l^\prime\,l}(T(t))
 +n_{l^\prime}(t)n_e(t)U_{l^\prime\,l}^{\star}(T(t)) \\ \nonumber
 &&+n_{l^\prime}(t)V_{l^\prime\,l}(T(t)) I(t)+n_{l^\prime}(t)n_e(t)V_{l^\prime\,l}^{\star}(T(t))I(t)] 
\\ \nonumber
&&-\sum_{l^\prime}n_l(t) n_e(t) C_{l\,l^\prime}(T(t))\\ \nonumber
&&-\sum_{l^\prime}\oint\,d\Omega\frac{d\nu}{h\nu}[
n_l(t) U_{l\,l^\prime}(T(t))
+n_l(t)n_e(t)U_{l\,l^\prime}^{\star}(T(t)) \\ \nonumber
&&+n_l(t) V_{l\,l^\prime}(T(t))I(t)
+n_l(t)n_e(t)V_{l\,l^\prime}^{\star}(T(t))I(t)] = \dot{n}_l \delta \, t.
\nonumber \\ \nonumber
\end{eqnarray}

Here $\dot{n}_{l,n-1}$ is the solution from the previous time step with time index $(n-1)$, $I(t)$ is an integral that depends nonlinearly on $n(t)$ and $T(t)$. 
The function $C_{l^\prime\, l}(T(t))$ is the temperature-dependent collision rate coefficient, and the functions $U_{l\,l^{\prime}}(t)$, $U_{l,l^{\prime}}^{\star}(t)$, $V_{l,l^{\prime}}(t)$ and 
$V_{l,l^{\prime}}^{\star}(t)$
are time-generalized versions of the $U$ and $V$ functions from \citet[][]{1992A&A...262..209R}, defined as follows.
For line transitions between $l$ and $l^{\prime}$ we have
\noindent
\begin{eqnarray}
&&U_{l,l^{\prime}}(\bm{\Omega},\nu,t) = \left\{
        \begin{array}{ll}
             & \frac{h\,\nu}{4\,\pi} A_{l,l^{\prime}} \varphi_{l,l^{\prime}}(\bm{\Omega},\nu,t), 
                 \quad l>l^{\prime},  \\ \nonumber
             & 0, \quad l<l^{\prime},
        \end{array}
    \right\},
\nonumber \\ \nonumber
&&U_{l,l^{\prime}}^{\star}(\bm{\Omega},\nu,t) = 0,
\nonumber \\ \nonumber
&&V_{l,l^{\prime}}(\bm{\Omega},\nu,t) = \frac{h\,\nu}{4\,\pi} B_{l,l^{\prime}} \varphi_{l,l^{\prime}}(\bm{\Omega},\nu,t),
\nonumber \\ \nonumber
&&V_{l,l^{\prime}}^{\star}(\bm{\Omega},\nu,t) = 0,
\nonumber \\ \nonumber
\end{eqnarray}
where $A_{l,l^{\prime}}$ and $B_{l,l^{\prime}}$ are the Einstein coefficients 
and $\varphi_{l,l^{\prime}}=\varphi_{l^{\prime}\,l}$ is the line profile function. 
Here $h$ and $k_b$ are Planck's constant and Boltzmann's constant, respectively. 
Here $V_{l,l^{\prime}}$ has the same form for both $l>l^{\prime}$ and $l<l^{\prime}$, 
although depending on whether it is an upward or a downward transition, we need to use appropriate Einstein $B$ coefficient.

Similarly for a continuum transition between $l$ and $l^{\prime}$, we have
\begin{eqnarray}
&&U_{l,l^{\prime}}(\nu,t) = 0,
\nonumber \\ \nonumber
&&U_{l,l^{\prime}}^{\star}(\nu,t) = 
\left\{
        \begin{array}{ll}
     &  \Phi_{l,l^{\prime}}(T(t))\frac{2\,h\,\nu^3}{C^2}
       e^{\frac{-h\,\nu}{k_b T(t)}} a_{l,l^{\prime}}(\nu), 
              \quad l>l^{\prime},  \\ \nonumber
     & 0, \quad l<l^{\prime},
        \end{array}
    \right\},
\nonumber \\ \nonumber
&&V_{l,l^{\prime}}(\nu,t) = 
\left\{
        \begin{array}{ll}
     & 0,   \quad l>l^{\prime},  \\ \nonumber
         
     & a_{l,l^{\prime}}(\nu), \quad l<l^{\prime},
        \end{array}
    \right\},
\nonumber \\ \nonumber
&&V_{l,l^{\prime}}^{\star}(\nu,t) = 
\left\{
        \begin{array}{ll}
     &  \Phi_{l,l^{\prime}}(T(t))\,e^{\frac{-h\,\nu}{k_b T(t)}} a_{l,l^{\prime}}(\nu), 
              \quad l>l^{\prime},  \\ \nonumber
     & 0, \quad l<l^{\prime},
        \end{array}
    \right\},
\nonumber \\ \nonumber
\end{eqnarray}
where $C$ is speed of light, and 
$a_{l,l^{\prime}}(\nu)$ is the photoionization crosssection, and $\Phi(T(t))$ is the Saha-Boltzmann factor, given by

\noindent
\begin{eqnarray}
&&\Phi(T(t)) = \frac{g_{l^{\prime}}}{2\,g_{l}} \left(\frac{h^2}{2\pi\,m_e\,k_b\,T(t)}\right)^{\frac{3}{2}}
\exp{\left[\frac{(E_l-E_l^{\prime})}{k_b\,T(t)}\right]}.
\nonumber \\ \nonumber
\end{eqnarray}
Thus, the time-dependent radiative rate coefficient $R_{l,l^{\prime}}$ is defined as

\noindent
\begin{eqnarray}
&&R_{l,l^{\prime}} = \int  d\,t \oint  d\,\bm{\Omega} \frac{ d\,\nu}{h\,\nu}
[(U_{l,l^{\prime}}(\bm{\Omega},\nu,t)+n_e(t) U_{l,l^{\prime}}^{\star}(\bm{\Omega},\nu,t))+
 (V_{l,l^{\prime}}(\bm{\Omega},\nu,t)+n_e(t) V_{l,l^{\prime}}^{\star}(\bm{\Omega},\nu,t)) I_{\bm{\Omega},\nu}(t)],
\nonumber \\ \nonumber
\end{eqnarray}
defined for both lines and continua and for transitions involving both $l>l^{\prime}$ 
and $l<l^{\prime}$.

First, we express all of the time-dependent quantities except $I(t)$ in the form of linear polynomials as follows,
\begin{eqnarray}
n_e(t)=n_e+t\,\dot{n}_e,\nonumber \\
n_l(t)=n_l+t\,\dot{n}_l, \nonumber \\
\varphi(t) = \varphi+t\,\dot{\varphi}, \nonumber \\
T(t)=T+t\,\dot{T}, \nonumber \\
\end{eqnarray}
and so on. For simplicity, the terms corresponding to zeroth order of 
$t$ (namely, $t^0$) are denoted by the same notation as the original time-dependent function.
For example, the $t^0$ term for temperature function $T(t)$ is denoted as $T$ itself. 
By substituting the above expansion in collisional and radiative rate integrals, we can separate the terms containing $t^0$ and $t^{(1)}$ in these rate equations.
We can rewrite the rate system as 

\noindent
\begin{eqnarray}
&&\sum_{l^\prime}\Big\{
(n_{l^\prime}n_e+t\,(\dot{n}_{l^\prime}n_e+\dot{n}_e n_{l^\prime})) C_{l^\prime\,l} \\ \nonumber
&&+(n_{l^\prime}\,n_e+t\,(\dot{n}_{l^\prime} n_e
+\dot{n}_e n_{l^\prime}))\,t\,\frac{\partial C_{l^\prime\,l}}{\partial T}\, \dot{T} \Big\} \\ \nonumber
&&+\sum_{l^{\prime}}\oint d\Omega \frac{d\nu}{h\nu}
\Big\{
n_{l^\prime} U_{l^\prime\,l}+t\,(\dot{n}_{l^\prime} U_{l^\prime\,l}+n_{l^{\prime}} 
\frac{\partial U_{l^\prime\,l}}{\partial T}\,\dot{T})\\ \nonumber
&&+(n_{l^\prime} n_e + t\, (\dot{n}_{l^\prime} n_e+\dot{n}_e n_{l^\prime}) ) U^{\star}_{l^\prime\,l}
+(n_{l^\prime} n_e + t\, (\dot{n}_{l^\prime} n_e+\dot{n}_e n_{l^\prime}) ) 
\,t\,\frac{\partial U^{\star}_{l^\prime\,l}}{\partial T}\, \dot{T} \Big\} \\ \nonumber
&&+\sum_{l^{\prime}}\oint d\Omega \frac{d\nu}{h\nu}
\Big\{
n_{l^\prime} V_{l^\prime\,l}+t\,(\dot{n}_{l^\prime} V_{l^\prime\,l}+n_{l^{\prime}} 
\frac{\partial V_{l^\prime\,l}}{\partial T} \dot{T}) \\ \nonumber
&&+(n_{l^\prime} n_e + t\, (\dot{n}_{l^\prime} n_e+\dot{n}_e n_{l^\prime}) ) V^{\star}_{l^\prime\,l}
+(n_{l^\prime} n_e + t\, (\dot{n}_{l^\prime} n_e+\dot{n}_e n_{l^\prime}) ) 
\,t\,\frac{\partial V^{\star}_{l^\prime\,l}}{\partial T} \dot{T} \Big\} I(t) 
\\ \nonumber
&&-\sum_{l^\prime}\,(n_l n_e+t\,(\dot{n}_l n_e+n_l \dot{n}_e) ) C_{l\,l^\prime} \\ \nonumber
&&-\sum_{l^\prime}\,(n_l n_e+t\,(\dot{n}_l n_e+n_l \dot{n}_e) ) \,t 
\frac{\partial C_{l\,l^\prime}}{\partial T} \dot{T} \\ \nonumber
&&-\sum_{l^\prime} \oint\,d\Omega\frac{d\nu}{h\nu}
\Big \{
n_l U_{l\,l^\prime} + t\, (\dot{n}_l U_{l\,l^\prime}+ n_l 
\frac{\partial U_{l\,l^\prime}}{\partial T}\,\dot{T}) \\ \nonumber
&&+(n_l n_e + t\, (\dot{n}_{l} n_e+\dot{n}_e n_{l}) ) U^{\star}_{l\,l^\prime}
+(n_{l} n_e + t\, (\dot{n}_{l} n_e+\dot{n}_e n_{l}) ) 
\,t\,\frac{\partial U^{\star}_{l\,l^\prime}}{\partial T}\, \dot{T} \Big\} \\ \nonumber
&&-\sum_{l^{\prime}}\oint d\Omega \frac{d\nu}{h\nu}
\Big\{
n_{l} V_{l\,l^\prime}+t\,(\dot{n}_{l} V_{l\,l^\prime}+n_{l} 
\frac{\partial V_{l\,l^\prime}}{\partial T} \dot{T}) \\ \nonumber
&&+(n_{l} n_e + t\, (\dot{n}_{l} n_e+\dot{n}_e n_{l}) ) V^{\star}_{l\,l^\prime}
+(n_{l} n_e + t\, (\dot{n}_{l} n_e+\dot{n}_e n_{l}) ) 
\,t\,\frac{\partial V^{\star}_{l\,l^\prime}}{\partial T} \dot{T} \Big\} I(t) 
\\ \nonumber
&&= \dot{n}_l \delta \,t.
\nonumber \\ \nonumber
\end{eqnarray}

Ignoring all terms that are second order or higher in time, we have

\noindent
\begin{eqnarray}
&&\sum_{l^\prime}\Big\{
(n_{l^\prime}n_e+t\,(\dot{n}_{l^\prime}n_e+\dot{n}_e n_{l^\prime})) C_{l^\prime\,l} \\ \nonumber
&&+(n_{l^\prime}\,n_e)
\,t\,\frac{\partial C_{l^\prime\,l}}{\partial T}\, \dot{T} \Big\} \\ \nonumber
&&+\sum_{l^{\prime}}\oint d\Omega \frac{d\nu}{h\nu}
\Big\{
n_{l^\prime} U_{l^\prime\,l}+t\,(\dot{n}_{l^\prime} U_{l^\prime\,l}+n_{l^{\prime}} 
\frac{\partial U_{l^\prime\,l}}{\partial T}\,\dot{T})\\ \nonumber
&&+(n_{l^\prime} n_e + t\, (\dot{n}_{l^\prime} n_e+\dot{n}_e n_{l^\prime}) ) U^{\star}_{l^\prime\,l}
+(n_{l^\prime} n_e ) 
\,t\,\frac{\partial U^{\star}_{l^\prime\,l}}{\partial T}\, \dot{T} \Big\} \\ \nonumber
&&+\sum_{l^{\prime}}\oint d\Omega \frac{d\nu}{h\nu}
\Big\{
n_{l^\prime} V_{l^\prime\,l}+t\,(\dot{n}_{l^\prime} V_{l^\prime\,l}+n_{l^{\prime}} 
\frac{\partial V_{l^\prime\,l}}{\partial T} \dot{T}) \\ \nonumber
&&+(n_{l^\prime} n_e + t\, (\dot{n}_{l^\prime} n_e+\dot{n}_e n_{l^\prime}) ) V^{\star}_{l^\prime\,l}
+(n_{l^\prime} n_e ) 
\,t\,\frac{\partial V^{\star}_{l^\prime\,l}}{\partial T} \dot{T} \Big\} I(t) 
\\ \nonumber
&&- \sum_{l^{\prime}} 
\Big\{
(n_{l}n_e+t\,(\dot{n}_{l}n_e+\dot{n}_e n_{l})) C_{l\,l^\prime} \\ \nonumber
&&+(n_{l}\,n_e)
\,t\,\frac{\partial C_{l\,l^\prime}}{\partial T}\, \dot{T} \Big\} \\ \nonumber
&&-\sum_{l^{\prime}}\oint d\Omega \frac{d\nu}{h\nu}
\Big\{
n_{l} U_{l\,l^\prime}+t\,(\dot{n}_{l} U_{l\,l^\prime}+n_{l} 
\frac{\partial U_{l\,l^\prime}}{\partial T}\,\dot{T})\\ \nonumber
&&+(n_{l} n_e + t\, (\dot{n}_{l} n_e+\dot{n}_e n_{l}) ) U^{\star}_{l\,l^\prime}
+(n_{l} n_e ) 
\,t\,\frac{\partial U^{\star}_{l\,l^\prime}}{\partial T}\, \dot{T} \Big\} \\ \nonumber
&&-\sum_{l^{\prime}}\oint d\Omega \frac{d\nu}{h\nu}
\Big\{
n_{l} V_{l\,l^\prime}+t\,(\dot{n}_{l} V_{l\,l^\prime}+n_{l} 
\frac{\partial V_{l\,l^\prime}}{\partial T} \dot{T}) \\ \nonumber
&&+(n_{l} n_e + t\, (\dot{n}_{l} n_e+\dot{n}_e n_{l}) ) V^{\star}_{l\,l^\prime}
+(n_{l} n_e ) 
\,t\,\frac{\partial V^{\star}_{l\,l^\prime}}{\partial T} \dot{T} \Big\} I(t) \\ \nonumber
&&=\dot{n}_l \delta \,t.
\nonumber
\end{eqnarray}

We now integrate with respect to time, frequency, and angle, and re-organize to obtain

\noindent
\begin{eqnarray}
&&\dot{n}_l\,\delta\,t - 
\Big\{\sum_{l^\prime} \dot{n}_{l^\prime} n_e \hat{D}_{l^\prime\,l}-
\sum_{l^\prime} \dot{n}_l n_e \bar{D}_{l\,l^\prime}\Big\} \\ \nonumber
&&- \Big\{\sum_{l^\prime}\dot{n}_{l^\prime}
\Big( \hat{U}_{l^\prime\,l}+n_e \hat{U}^{\star}_{l^\prime\,l}+\hat{J}_{l^\prime\,l}
+n_e \hat{J}^{\star}_{l^\prime\,l}\Big)\\ \nonumber
&&-\sum_{l^\prime} \dot{n}_l \Big(\hat{U}_{l\,l^\prime}+n_e \hat{U}^{\star}_{l\,l^\prime}
+\hat{J}_{l\,l^\prime}+n_e \hat{J}^{\star}_{l\,l^\prime} 
\Big)\Big\}\\ \nonumber
&&-\dot{n}_e\Big\{
\sum_{l^\prime} \Big(n_{l^\prime} \hat{D}_{l^\prime\,l}-n_l \hat{D}_{l\,l^\prime}\Big)
+\sum_{l^\prime} n_{l^\prime}\Big(\hat{U}^{\star}_{l^\prime\,l}
+\hat{J}^{\star}_{l^\prime\,l} \Big) 
-n_l \Big( \hat{U}^{\star}_{l\,l^\prime}+\hat{J}^{\star}_{l\,l^\prime}\Big) 
\Big \}\\ \nonumber
&&-\dot{T} \Big \{
\sum_{l^\prime} \Big(n_{l^\prime} n_e \hat{d\,D}_{l^\prime\,l}-n_l n_e \hat{d\,D}_{l\,l^\prime}\Big) \\ \nonumber
&&+\sum_{l^\prime} n_{l^\prime}\Big(\hat{d\,U}_{l^\prime\,l}+n_e \hat{d\,U}^{\star}_{l^\prime\,l}
+\hat{d\,J}_{l^\prime\,l}+n_e \hat{d\,J}^{\star}_{l^\prime\,l} \Big) \\ \nonumber
&&-n_l \Big( \hat{d\,U}_{l\,l^\prime}+n_e \hat{d\,U}^{\star}_{l\,l^\prime}
+\hat{d\,J}_{l\,l^\prime}+n_e \hat{d\,J}^{\star}_{l\,l^\prime}\Big) 
\Big \}
=  \\ \nonumber
&&\sum_{l^\prime}\Big(n_{l^\prime} n_e \bar{D}_{l^\prime\,l}-n_l n_e \bar{D}_{l\,l^\prime}\Big) \\ \nonumber
&&+\Big\{
\sum_{l^\prime} \Big(
n_{l^\prime} \bar{U}_{l^\prime\,l}+n_{l^\prime} n_e \bar{U}^{\star}_{l^\prime\,l}
+n_{l^\prime} \bar{J}_{l^\prime\,l}+n_{l^\prime} n_e \bar{J}^{\star}_{l^\prime\,l}
\Big)\\ \nonumber
&&-\Big( 
n_{l} \bar{U}_{l\,l^\prime}+n_{l} n_e \bar{U}^{\star}_{l\,l^\prime}
+n_{l} \bar{J}_{l\,l^\prime}+n_{l} n_e \bar{J}^{\star}_{l\,l^\prime}
\Big),
\label{rate-system-integrated} 
\end{eqnarray}
where various integrals involving the collisional and the radiative rates, denoted as 
$\bar{D}$, $\hat{D}$, $\hat{d\,D}$, $\bar{U}$, $\bar{U}^{\star}$, 
$\hat{U}$, $\hat{U}^{\star}$, $\hat{d\,U}$, $\hat{d\,U}^{\star}$
$\bar{J}$, $\bar{J}^{\star}$, $\hat{J}$, $\hat{J}^{\star}$, $\hat{d\,J}$ and $\hat{d\,J}^{\star}$, are defined as

\begin{eqnarray}
&& \bar{D}_{l\,l^{\prime}}=\int_{0}^{\delta\,t}  d\,t\, C_{l\,l^{\prime}}, \\ \nonumber 
&& \hat{D}_{l\,l^{\prime}}=\int_{0}^{\delta\,t} \,t\,  d\,t\, C_{l\,l^{\prime}}, \\ \nonumber
&& \hat{d\,D}_{l\,l^{\prime}}=\int_{0}^{\delta\,t} \,t\,  d\,t\, 
\frac{\partial C_{l\,l^{\prime}}}{\partial T}, \\ \nonumber
\end{eqnarray}

\begin{eqnarray}
&& \bar{U}_{l\,l^{\prime}}=\int_{0}^{\delta\,t}  d\,t\, U_{l\,l^{\prime}}, \\ \nonumber 
&& \hat{U}_{l\,l^{\prime}}=\int_{0}^{\delta\,t} \,t\,  d\,t\, U_{l\,l^{\prime}}, 
\\ \nonumber
&& \hat{d\,U}_{l\,l^{\prime}}=\int_{0}^{\delta\,t} \,t\,  d\,t\, 
\frac{\partial U_{l\,l^{\prime}}}{\partial T}, \\ \nonumber
\end{eqnarray}

\begin{eqnarray}
&& \bar{U}^{\star}_{l\,l^{\prime}}=\int_{0}^{\delta\,t}  d\,t\, U^{\star}_{l\,l^{\prime}}, \\ \nonumber 
&& \hat{U}^{\star}_{l\,l^{\prime}}=\int_{0}^{\delta\,t} \,t\,  d\,t\, 
U^{\star}_{l\,l^{\prime}}, \\ \nonumber
&& \hat{d\,U}^{\star}_{l\,l^{\prime}}=\int_{0}^{\delta\,t} \,t\,  d\,t\, 
\frac{\partial U^{\star}_{l\,l^{\prime}}}{\partial T}, \\ \nonumber
\end{eqnarray}

\begin{eqnarray}
&& \bar{J}_{l\,l^{\prime}}=\int_{0}^{\delta\,t}  d\,t\, V_{l\,l^{\prime}} I(t), \\ \nonumber 
&& \hat{J}_{l\,l^{\prime}}=\int_{0}^{\delta\,t} \,t\,  d\,t\, V_{l\,l^{\prime}} I(t), \\ \nonumber
&& \hat{d\,J}_{l\,l^{\prime}}=\int_{0}^{\delta\,t} \,t\,  d\,t\, 
\frac{\partial V_{l\,l^{\prime}}}{\partial T} I(t), \\ \nonumber
\end{eqnarray}

\begin{eqnarray}
&& \bar{J}^{\star}_{l\,l^{\prime}}=\int_{0}^{\delta\,t}  d\,t\, V^{\star}_{l\,l^{\prime}} I(t), 
\\ \nonumber 
&& \hat{J}^{\star}_{l\,l^{\prime}}=\int_{0}^{\delta\,t} \,t\,  d\,t\, V^{\star}_{l\,l^{\prime}} I(t) ,
\\ \nonumber
&& \hat{d\,J}^{\star}_{l\,l^{\prime}}=\int_{0}^{\delta\,t} \,t\,  d\,t\, 
\frac{\partial V^{\star}_{l\,l^{\prime}}}{\partial T} I(t). \\ \nonumber
\end{eqnarray}

\section{Acceleration: Time-dependent Preconditioned MALI Scheme}
\label{appendixb}
We follow the preconditioning approach by \citet[][]{1992A&A...262..209R} to treat nonlinearities in the rate system and generalize the method for the time-dependent case. Applying operator splitting on the radiation field leads to
\begin{equation}
I_{\bm{\Omega},\nu,t}=\Psi^{\star}_{\bm{\Omega},\nu,t}\left[t\,\dot{\eta}_{\bm{\Omega},\nu}\right]+
(\Psi_{\bm{\Omega},\nu,t}-\Psi^{\star}_{\bm{\Omega},\nu,t})\left[t\,\dot{\eta}^{\dagger}_{\bm{\Omega},\nu}\right].
\end{equation}
Integrating over time, we have 
\begin{equation}
\bar{I}_{\bm{\Omega},\nu}=\bar{\Psi}^{\star}_{\bm{\Omega},\nu}\left[\dot{\eta}_{\bm{\Omega},\nu}\right]+
(\bar{\Psi}_{\bm{\Omega},\nu}-\bar{\Psi}^{\star}_{\bm{\Omega},\nu})\left[\dot{\eta}^{\dagger}_{\bm{\Omega},\nu}\right],
\end{equation}
and
\begin{equation}
\hat{I}_{\bm{\Omega},\nu}=\hat{\Psi}^{\star}_{\bm{\Omega},\nu}\left[\dot{\eta}_{\bm{\Omega},\nu}\right]+
(\hat{\Psi}_{\bm{\Omega},\nu}-\hat{\Psi}^{\star}_{\bm{\Omega},\nu})\left[\dot{\eta}^{\dagger}_{\bm{\Omega},\nu}\right],
\end{equation}
where
\begin{equation}
\dot{\eta}_{\bm{\Omega},\nu}
=\sum_{l>l^{\prime}} \dot{n}_l 
\left[ U_{l,l^{\prime}} + n_e\, U^{\star}_{l,l^{\prime}} \right]
+\dot{T} \sum_{l>l^{\prime}} n_l 
\left[ \frac{\partial U_{l,l^{\prime}}}{\partial T} + n_e\, \frac{\partial U_{l,l^{\prime}}^{\star}}{\partial T} \right]+\dot{n}_e \sum_{l>l^{\prime}} n_l\, U^{\star}_{l,l^{\prime}}.
\end{equation}
Substituting these expressions in Equation~(\ref{rate-system-integrated}) and simplifying, we obtain 

\begin{eqnarray}\label{rate-system-integrated-acc} 
&& \dot{n}_l\,\delta\,t - 
\Big\{\sum_{l^\prime} \dot{n}_{l^\prime} n_e \hat{D}_{l^\prime\,l}-
\sum_{l^\prime} \dot{n}_l n_e \bar{D}_{l\,l^\prime}\Big\} \\ \nonumber 
&&-\Big\{\sum_{l^\prime}\dot{n}_{l^\prime}
\Big( \hat{U}_{l^\prime\,l}+ n_e \hat{U}^{\star}_{l^\prime\,l}
+\hat{J}_{l^\prime\,l}+ n_e \hat{J}^{\star}_{l^\prime\,l}\Big)\\ \nonumber
&&-\sum_{l^\prime} \dot{n}_l \Big(\hat{U}_{l\,l^\prime}+n_e \hat{U}^{\star}_{l\,l^\prime}
+\hat{J}_{l\,l^\prime}+n_e \hat{J}^{\star}_{l\,l^\prime} 
\Big)\Big\}\\ \nonumber
&&-\dot{n}_e\Big\{
\sum_{l^\prime} \Big(n_{l^\prime} \hat{D}_{l^\prime\,l}-n_l \hat{D}_{l\,l^\prime}\Big)
+\sum_{l^\prime} n_{l^\prime}\Big(\hat{U}^{\star}_{l^\prime\,l}+\hat{J}^{\star}_{l^\prime\,l} \Big) 
-n_l \Big( \hat{U}^{\star}_{l\,l^\prime}+\hat{J}^{\star}_{l\,l^\prime}\Big) 
\Big \}\\ \nonumber
&&-\dot{T} \Big \{
\sum_{l^\prime} \Big(n_{l^\prime} n_e \hat{d\,D}_{l^\prime\,l}-n_l n_e \hat{d\,D}_{l\,l^\prime}\Big) \\ \nonumber
&&+\sum_{l^\prime} n_{l^\prime}\Big(\hat{d\,U}_{l^\prime\,l}+n_e \hat{d\,U}^{\star}_{l^\prime\,l}
+\hat{d\,J}_{l^\prime\,l}+n_e \hat{d\,J}^{\star}_{l^\prime\,l} \Big) \\ \nonumber
&&-n_l \Big( \hat{d\,U}_{l\,l^\prime}+n_e \hat{d\,U}^{\star}_{l\,l^\prime}
+\hat{d\,J}_{l\,l^\prime}+n_e \hat{d\,J}^{\star}_{l\,l^\prime}\Big) 
\Big \}\\ \nonumber
&&+
\sum_{l^{\prime}} \dot{n}_{l^{\prime}} 
\Big(\hat{\Psi}^{\star}_{I,l^{\prime},l} 
+\hat{\Psi}^{\star}_{II,l^{\prime},l} 
+\hat{\Psi}^{\star}_{III,l^{\prime},l}
\Big)
-\Big(\sum_{l^{\prime}} \dot{n}_l \hat{\Psi}^{\star}_{Ia,l,l^{\prime}}
+\dot{n}_e \hat{\Psi}^{\star}_{IIa,l,l^{\prime}}
+\dot{T}\,\, \hat{\Psi}^{\star}_{IIIa,l,l^{\prime}} \Big)\\ \nonumber 
&&=\\ \nonumber 
&&\sum_{l^\prime}\Big(n_{l^\prime} n_e \bar{D}_{l^\prime\,l}-n_l n_e \bar{D}_{l\,l^\prime}\Big) \\ \nonumber
&&+\sum_{l^\prime} \Big(
n_{l^\prime} \bar{U}_{l^\prime\,l}+n_{l^\prime} n_e \bar{U}^{\star}_{l^\prime\,l}
+n_{l^\prime} \bar{J}_{l^\prime\,l}+n_{l^\prime} n_e \bar{J}^{\star}_{l^\prime\,l}
\Big)\\ \nonumber
&&-\Big( 
n_{l} \bar{U}_{l\,l^\prime}+n_{l} n_e \bar{U}^{\star}_{l\,l^\prime}
+ n_{l} \bar{J}_{l\,l^\prime}+n_{l} n_e \bar{J}^{\star}_{l\,l^\prime}
\Big) \\ \nonumber
&&+ \bar{\Psi}^{\star}_{IV,l} .
\end{eqnarray}

where
\begin{eqnarray}\label{psiI}
&&\hat{\Psi}^{\star}_{I,l^{\prime},l}=\Bigg[ 
\left(V_{l^{\prime},l}+n_e V^{\star}_{l^{\prime},l}\right)
\hat{\Psi}^{\star} \\ \nonumber
&& \times \left(\sum_{m,m^{\prime}} \dot{n}_m^{\dagger} \left(U_{m,m^{\prime}}+n_e U^{\star}_{m,m^{\prime}}\right)
+ n_m \dot{n}_e^{\dagger} U_{m,m^{\prime}}^{\star}
+ n_m  \dot{T}^{\dagger}\Big(\frac{\partial U_{m,m^{\prime}}}{\partial T}+n_e \frac{\partial U_{m,m^{\prime}}^{\star}}{\partial T} \Big)
\right)\\ \nonumber 
&&+\left(\sum_{m} n_m \left(V_{m,l}+n_e V_{m,l}^{\star}\right)-
n_l \left(V_{l,m}+n_e V_{l,m}^{\star}\right) \right) \bar{\Psi}^{\star} 
\left(\sum_{m^{\prime}} U_{l^{\prime},m^{\prime}}+n_e U^{\star}_{l^{\prime},m^{\prime}} \right)\nonumber \\
&&-\left(\sum_{m}\dot{n}_m^{\dagger} 
\left(V_{m,l}+n_e V_{m,l}^{\star}\right)-\dot{n}_l^{\dagger} 
\left(V_{l,m}+n_e V_{l,m}^{\star}\right) \right) \hat{\Psi}^{\star} 
\left(\sum_{m^{\prime}} U_{l^{\prime},m^{\prime}}+n_e U^{\star}_{l^{\prime},m^{\prime}} \right)
\Bigg], \\ \nonumber
\end{eqnarray}

\begin{eqnarray}\label{PsiII}
&&\hat{\Psi}^{\star}_{II,l^{\prime},l}=
\dot{n}^{\dagger}_e \Big(\sum_{m} n_l V^{\star}_{l,m}-n_m V^{\star}_{m,l}\Big)\hat{\Psi}^{\star}\Big( \sum_{m^{\prime}} U_{l^{\prime},m^{\prime}}+n_e U_{l^{\prime},m^{\prime}}^{\star}\Big),\\ \nonumber 
\end{eqnarray}

\begin{eqnarray}\label{PsiIII}
&&\hat{\Psi}^{\star}_{III,l^{\prime},l}=\dot{T}^{\dagger}
\Big(\sum_{m} n_l \Big( \frac{\partial V_{l,m}}{\partial T}+
n_e \frac{\partial V^{\star}_{l,m}}{\partial T}\Big)\\ \nonumber
&&-n_m \Big(\frac{\partial V_{m,l}}{\partial T}+ n_e \frac{\partial V^{\star}_{m,l}}{\partial T}\Big)\Big)
\hat{\Psi}^{\star}\Big( \sum_{m^{\prime}} U_{l^{\prime},m^{\prime}} +n_e U^{\star}_{l^{\prime},m^{\prime}}\Big),
\\ \nonumber 
\end{eqnarray}

\begin{eqnarray}\label{PsiIa}
&&\hat{\Psi}^{\star}_{Ia,l,l^{\prime}}=\left( V_{l,l^{\prime}}+n_e V^{\star}_{l,l^{\prime}}\right) \hat{\Psi}^{\star} \\ \nonumber
&& \times \left(\sum_{m,m^{\prime}}\dot{n}_m^{\dagger} \left(U_{m,m^{\prime}}+n_e U^{\star}_{m,m^{\prime}} \right) 
+ n_m \dot{n}_e^{\dagger} U_{m,m^{\prime}}^{\star}
+ n_m  \dot{T}^{\dagger}\Big(\frac{\partial U_{m,m^{\prime}}}{\partial T}+n_e \frac{\partial U_{m,m^{\prime}}^{\star}}{\partial T} \Big)
\right),
\end{eqnarray}

\begin{eqnarray}\label{PsiIIa}
&&\hat{\Psi}^{\star}_{IIa,l,l^{\prime}}=
\Bigg \{ 
 \sum_{l^\prime} \Big(n_l V^{\star}_{l,l^{\prime}}-n_{l^\prime} V^{\star}_{l^{\prime},l}\Big)\hat{\Psi}^{\star}
\Big( \sum_{m,m^{\prime}} \dot{n}_m^{\dagger}\Big(U_{m,m^{\prime}}+n_e U_{m,m^{\prime}}^{\star}\Big) \\ \nonumber
&&+ n_m \dot{T}^{\dagger} \Big(\frac{\partial U_{m,m^{\prime}}}{\partial T}+n_e \frac{\partial U_{m,m^{\prime}}^{\star}}{\partial T}  \Big) \Big )\\ \nonumber
&&-\sum_{l^\prime} \Big(n_l \,\Big(\frac{\partial V_{l,l^{\prime}}}{\partial T}+n_e \,\frac{\partial V^{\star}_{l,l^{\prime}} }{\partial T}\Big)-n_{l^\prime} \Big(\frac{\partial V_{l^{\prime},l}}{\partial T}+n_e\, \frac {\partial V^{\star}_{l^{\prime},l}}{\partial T}\Big)\Big)\hat{\Psi}^{\star} \Big( \sum_{m,m^{\prime}}
n_m \, U_{m,m^{\prime}}^{\star} \dot{T}^{\dagger}
 \Big)\\ \nonumber
&&-\sum_{l^\prime} \Big(n_l (V_{l,l^{\prime}}+n_e \,V^{\star}_{l,l^{\prime}})-n_{l^\prime} (V_{l^{\prime},l}+n_e\, V^{\star}_{l^{\prime},l})\Big)\bar{\Psi}^{\star}\Big( \sum_{m,m^{\prime}} n_m \, U_{m,m^{\prime}}^{\star} \Big)\\ \nonumber
&&-\sum_{l^\prime} \Big(\dot{n}^{\dagger}_l (V_{l,l^{\prime}}+n_e \,V^{\star}_{l,l^{\prime}})-\dot{n}^{\dagger}_{l^\prime} (V_{l^{\prime},l}+n_e\, V^{\star}_{l^{\prime},l})\Big)\hat{\Psi}^{\star}\Big( \sum_{m,m^{\prime}} n_m \, U_{m,m^{\prime}}^{\star} \Big)\Bigg \}, \\ \nonumber
\end{eqnarray}

\begin{eqnarray}\label{PsiIIIa}
&&\hat{\Psi}^{\star}_{IIIa,l,l^{\prime}}=\Bigg \{ \sum_{l^{\prime}}  
\Big( n_l \Big( \frac{\partial V_{l,l^{\prime}}}{\partial T}+
n_e \frac{\partial V^{\star}_{l,l^{\prime}}}{\partial T}\Big)-n_{l^{\prime}} \Big(\frac{\partial V_{l^{\prime},l}}{\partial T}+ n_e \frac{\partial V^{\star}_{l^{\prime},l}}{\partial T}\Big)\Big)\\\nonumber
&&\times \hat{\Psi}^{\star}
\Big( \sum_{m,m^{\prime}}\dot{n}_m^{\dagger} \Big(U_{m,m^{\prime}} + n_e U^{\star}_{m,m^{\prime}}\Big)+ n_m \dot{n}_e^{\dagger} U^{\star}_{m,m^{\prime}} \Big) \\ \nonumber
&& -\sum_{l^{\prime}}  
\Big( n_l  V^{\star}_{l,l^{\prime}}-n_{l^{\prime}}   V^{\star}_{l^{\prime},l}\Big)
\hat{\Psi}^{\star}\Big( \sum_{m,m^{\prime}}
n_m \dot{n}_e^{\dagger} \Big(\frac{\partial U_{m,m^{\prime}}}{\partial T}+n_e \frac{\partial U_{m,m^{\prime}}^{\star}}{\partial T} \Big)\Big) \\ \nonumber
&&-\sum_{l^{\prime}}  
\Big( n_l  \Big( V_{l,l^{\prime}}+n_e V^{\star}_{l,l^{\prime}}\Big)-n_{l^{\prime}}  \Big( V_{l^{\prime},l} + n_e V^{\star}_{l^{\prime},l} \Big) \Big)
\bar{\Psi}^{\star}\Big( \sum_{m,m^{\prime}}
n_m  \Big(\frac{\partial U_{m,m^{\prime}}}{\partial T}+n_e \frac{\partial U_{m,m^{\prime}}^{\star}}{\partial T} \Big)\Big)\\\nonumber
&&-\sum_{l^{\prime}}  
\Big(\dot{n}^{\dagger}_l  \Big( V_{l,l^{\prime}}+n_e V^{\star}_{l,l^{\prime}}\Big) -\dot{n}^{\dagger}_{l^{\prime}}  \Big( V_{l^{\prime},l} + n_e V^{\star}_{l^{\prime},l} \Big) \Big)
\hat{\Psi}^{\star}\Big( \sum_{m,m^{\prime}}
n_m  \Big(\frac{\partial U_{m,m^{\prime}}}{\partial T}+n_e \frac{\partial U_{m,m^{\prime}}^{\star}}{\partial T} \Big)\Big)
\Bigg \}, \\ \nonumber
\end{eqnarray}

\begin{eqnarray}\label{PsiIV}
&&\bar{\Psi}^{\star}_{IV,l}=
\Big( \sum_{l^{\prime}} n_l 
\Big(V_{l,l^{\prime}}+n_e V^{\star}_{l,l^{\prime}} \Big)
-n_{l^{\prime}} \Big( V_{l^{\prime},l}+n_e V^{\star}_{l^{\prime},l}\Big) \Big)
\bar{\Psi}^{\star} \\ \nonumber 
&& \times \Big(\sum_{m,m^{\prime}}   \dot{n}^{\dagger}_m \Big(U_{m,m^{\prime}} + n_e U^{\star}_{m,m^{\prime}} \Big)
+ n_m \dot{n}_e^{\dagger} U_{m,m^{\prime}}^{\star} 
+ n_m  \dot{T}^{\dagger}\Big(\frac{\partial U_{m,m^{\prime}}}{\partial T}+n_e \frac{\partial U_{m,m^{\prime}}^{\star}}{\partial T} \Big)
\Big).
\end{eqnarray}

\bibliography{ms}{}
\bibliographystyle{aasjournal}

\end{document}